\input harvmac
\input amssym
\input epsf
\input miniltx.tex
\input graphicx.sty
\resetatcatcode

\def\unit{\relax{\rm 1\kern-.26em I}}
\def\nada{\relax{\rm 0\kern-.30em l}}
\def\tilde{\widetilde}

%\draftmode

%\def\Omega{\rho,\sigma,\nu  }

%% MACROS
\noblackbox
\def\IL{\relax{\rm I\kern-.18em L}}
\def\IH{\relax{\rm I\kern-.18em H}}
\def\IR{\relax{\rm I\kern-.18em R}}
\def\IC{\relax\hbox{$\inbar\kern-.3em{\rm C}$}}
\def\IZ{\relax\ifmmode\mathchoice
{\hbox{\cmss Z\kern-.4em Z}}{\hbox{\cmss Z\kern-.4em Z}}
{\lower.9pt\hbox{\cmsss Z\kern-.4em Z}} {\lower1.2pt\hbox{\cmsss
Z\kern-.4em Z}}\else{\cmss Z\kern-.4em Z}\fi}

\def\CN {{\cal N}}

\def\CF {{\cal F}}
\def\CJ {{\cal J}}

\def\CL {{\cal L}}

\def\CO {{\cal O}}

%% MORE MACROS

\def\CN {{\cal N}}

\def\CO {{\cal O}}

\def\Tr{{\rm Tr}}

\font\manual=manfnt \def\dbend{\lower3.5pt\hbox{\manual\char127}}

\def\IZ{\relax\ifmmode\mathchoice
{\hbox{\cmss Z\kern-.4em Z}}{\hbox{\cmss Z\kern-.4em Z}}
{\lower.9pt\hbox{\cmsss Z\kern-.4em Z}} {\lower1.2pt\hbox{\cmsss
Z\kern-.4em Z}}\else{\cmss Z\kern-.4em Z}\fi}

\def\bar{\overline}

\def\rt2{\sqrt{2}}
\def\irt2{{1\over\sqrt{2}}}

\def\hat{\widehat}
%  \slashchar puts a slash through a character to represent contraction
%  with Dirac matrices. Use \not instead for negation of relations, and use
%  \hbar for hbar.
\def\slashchar#1{\setbox0=\hbox{$#1$}           % set a box for #1
   \dimen0=\wd0                                 % and get its size
   \setbox1=\hbox{/} \dimen1=\wd1               % get size of /
   \ifdim\dimen0>\dimen1                        % #1 is bigger
      \rlap{\hbox to \dimen0{\hfil/\hfil}}      % so center / in box
      #1                                        % and print #1
   \else                                        % / is bigger
      \rlap{\hbox to \dimen1{\hfil$#1$\hfil}}   % so center #1
      /                                         % and print /
   \fi}

\def\foursqr#1#2{{\vcenter{\vbox{
    \hrule height.#2pt
    \hbox{\vrule width.#2pt height#1pt \kern#1pt
    \vrule width.#2pt}
    \hrule height.#2pt
    \hrule height.#2pt
    \hbox{\vrule width.#2pt height#1pt \kern#1pt
    \vrule width.#2pt}
    \hrule height.#2pt
        \hrule height.#2pt
    \hbox{\vrule width.#2pt height#1pt \kern#1pt
    \vrule width.#2pt}
    \hrule height.#2pt
        \hrule height.#2pt
    \hbox{\vrule width.#2pt height#1pt \kern#1pt
    \vrule width.#2pt}
    \hrule height.#2pt}}}}
\def\psqr#1#2{{\vcenter{\vbox{\hrule height.#2pt
    \hbox{\vrule width.#2pt height#1pt \kern#1pt
    \vrule width.#2pt}
    \hrule height.#2pt \hrule height.#2pt
    \hbox{\vrule width.#2pt height#1pt \kern#1pt
    \vrule width.#2pt}
    \hrule height.#2pt}}}}
\def\sqr#1#2{{\vcenter{\vbox{\hrule height.#2pt
    \hbox{\vrule width.#2pt height#1pt \kern#1pt
    \vrule width.#2pt}
    \hrule height.#2pt}}}}

\def\figin{\epsfcheck\figin}\def\figins{\epsfcheck\figins}
\def\epsfcheck{\ifx\epsfbox\UnDeFiNeD
\message{(NO epsf.tex, FIGURES WILL BE IGNORED)}
\gdef\figin##1{\vskip2in}\gdef\figins##1{\hskip.5in}% blank space instead
\else\message{(FIGURES WILL BE INCLUDED)}%
\gdef\figin##1{##1}\gdef\figins##1{##1}\fi}
\def\DefWarn#1{}
\def\figinsert{\goodbreak\midinsert}
\def\ifig#1#2#3{\DefWarn#1\xdef#1{fig.~\the\figno}
\writedef{#1\leftbracket fig.\noexpand~\the\figno}%
\figinsert\figin{\centerline{#3}}\medskip\centerline{\vbox{\baselineskip12pt
\advance\hsize by -1truein\noindent\footnotefont{\bf
Fig.~\the\figno:\ } \it#2}}
\bigskip\endinsert\global\advance\figno by1}

%\DorigoniRA
\lref\DorigoniRA{
  D.~Dorigoni, V.~S.~Rychkov,
  ``Scale Invariance + Unitarity $=>$ Conformal Invariance?,''
[arXiv:0910.1087 [hep-th]].
%%CITATION = arXiv:0910.1087%%
}

%\HeckmanQV
\lref\HeckmanQV{
  J.~J.~Heckman, Y.~Tachikawa, C.~Vafa and B.~Wecht,
  ``N = 1 SCFTs from Brane Monodromy,''
JHEP {\bf 1011}, 132 (2010).
[arXiv:1009.0017 [hep-th]].
%%CITATION = arXiv:1009.0017%%
}

%\IntriligatorJJ
\lref\IntriligatorJJ{
  K.~A.~Intriligator, B.~Wecht,
  ``The Exact superconformal R symmetry maximizes a,''
Nucl.\ Phys.\  {\bf B667}, 183-200 (2003).
[hep-th/0304128].
%%CITATION = hep-th/0304128%%
}

%\RattazziGJ
\lref\RattazziGJ{
  R.~Rattazzi, S.~Rychkov and A.~Vichi,
  ``Central Charge Bounds in 4D Conformal Field Theory,''
Phys.\ Rev.\ D {\bf 83}, 046011 (2011).
[arXiv:1009.2725 [hep-th]].
%%CITATION = arXiv:1009.2725%%
}

%\KomargodskiRB
\lref\KomargodskiRB{
  Z.~Komargodski, N.~Seiberg,
  ``Comments on Supercurrent Multiplets, Supersymmetric Field Theories and Supergravity,''
JHEP {\bf 1007}, 017 (2010).
[arXiv:1002.2228 [hep-th]].
%%CITATION = arXiv:1002.2228%%
}

%\AbelWV
\lref\AbelWV{
  S.~Abel, M.~Buican, Z.~Komargodski,
  ``Mapping Anomalous Currents in Supersymmetric Dualities,''
[arXiv:1105.2885 [hep-th]].
%%CITATION = CERN-PH-TH-2011-112%%
}

%\PolandWG
\lref\PolandWG{
  D.~Poland, D.~Simmons-Duffin,
  ``Bounds on 4D Conformal and Superconformal Field Theories,''
JHEP {\bf 1105}, 017 (2011).
[arXiv:1009.2087 [hep-th]].
%%CITATION = arXiv:1009.2087%%
}

%\HofmanAR
\lref\HofmanAR{
  D.~M.~Hofman, J.~Maldacena,
  ``Conformal collider physics: Energy and charge correlations,''
JHEP {\bf 0805}, 012 (2008).
[arXiv:0803.1467 [hep-th]].
%%CITATION = arXiv:0803.1467%%
}

%\BuicanEC
\lref\BuicanEC{
  M.~Buican,
  ``Non-Perturbative Constraints on Light Sparticles from Properties of the RG Flow,''
[arXiv:1206.3033 [hep-th]].
%%CITATION = arXiv:1206.3033%%
}

%\MaldacenaJN
\lref\MaldacenaJN{
  J.~Maldacena and A.~Zhiboedov,
  ``Constraining Conformal Field Theories with A Higher Spin Symmetry,''
J.\ Phys.\ A {\bf 46}, 214011 (2013).
[arXiv:1112.1016 [hep-th]].
%%CITATION = arXiv:1112.1016%%
}

%\BuicanTY
\lref\BuicanTY{
  M.~Buican,
  ``A Conjectured Bound on Accidental Symmetries,''
Phys.\ Rev.\ D {\bf 85}, 025020 (2012).
[arXiv:1109.3279 [hep-th]].
%%CITATION = CERN-PH-TH-2011-223%%
}

%\AntoniadisNJ
\lref\AntoniadisNJ{
  I.~Antoniadis and M.~Buican,
  ``Goldstinos, Supercurrents and Metastable SUSY Breaking in N=2 Supersymmetric Gauge Theories,''
JHEP {\bf 1104}, 101 (2011).
[arXiv:1005.3012 [hep-th]].
%%CITATION = arXiv:1005.3012%%
}

%\AntoniadisGN
\lref\AntoniadisGN{
  I.~Antoniadis, M.~Buican,
  ``On R-symmetric Fixed Points and Superconformality,''
Phys.\ Rev.\  {\bf D83}, 105011 (2011).
[arXiv:1102.2294 [hep-th]].
%%CITATION = arXiv:1102.2294%%
}

%\ZamolodchikovGT
\lref\ZamolodchikovGT{
  A.~B.~Zamolodchikov,
  ``Irreversibility of the Flux of the Renormalization Group in a 2D Field Theory,''
JETP Lett.\  {\bf 43}, 730-732 (1986).
}

%\PolchinskiDY
\lref\PolchinskiDY{
  J.~Polchinski,
  ``Scale And Conformal Invariance In Quantum Field Theory,''
Nucl.\ Phys.\  {\bf B303}, 226 (1988).
%%CITATION = UTTG-22-87%%
}

%\CardyCWA
\lref\CardyCWA{
  J.~L.~Cardy,
  ``Is There a c Theorem in Four-Dimensions?,''
Phys.\ Lett.\  {\bf B215}, 749-752 (1988).
}

%\KomargodskiVJ
\lref\KomargodskiVJ{
  Z.~Komargodski and A.~Schwimmer,
  ``On Renormalization Group Flows in Four Dimensions,''
  arXiv:1107.3987 [hep-th].
  %%CITATION = ARXIV:1107.3987;%%
}

%\AmaritiWC
\lref\AmaritiWC{
  A.~Amariti and K.~Intriligator,
  ``(Delta a) curiosities in some 4d susy RG flows,''
JHEP {\bf 1211}, 108 (2012).
[arXiv:1209.4311 [hep-th]].
%%CITATION = arXiv:1209.4311%%
}

%\ArgyresCN
\lref\ArgyresCN{
  P.~C.~Argyres and N.~Seiberg,
  ``S-duality in N=2 supersymmetric gauge theories,''
JHEP {\bf 0712}, 088 (2007).
[arXiv:0711.0054 [hep-th]].
%%CITATION = arXiv:0711.0054%%
}

%\ArgyresJJ
\lref\ArgyresJJ{
  P.~C.~Argyres and M.~R.~Douglas,
  ``New phenomena in SU(3) supersymmetric gauge theory,''
Nucl.\ Phys.\ B {\bf 448}, 93 (1995).
[hep-th/9505062].
%%CITATION = hep-th/9505062%%
}

%\ArgyresTQ
\lref\ArgyresTQ{
  P.~C.~Argyres and J.~R.~Wittig,
  ``Infinite coupling duals of N=2 gauge theories and new rank 1 superconformal field theories,''
JHEP {\bf 0801}, 074 (2008).
[arXiv:0712.2028 [hep-th]].
%%CITATION = arXiv:0712.2028%%
}

%\EguchiVU
\lref\EguchiVU{
  T.~Eguchi, K.~Hori, K.~Ito and S.~-K.~Yang,
  ``Study of N=2 superconformal field theories in four-dimensions,''
Nucl.\ Phys.\ B {\bf 471}, 430 (1996).
[hep-th/9603002].
%%CITATION = hep-th/9603002%%
}

%\DymarskyPQA
\lref\DymarskyPQA{
  A.~Dymarsky, Z.~Komargodski, A.~Schwimmer and S.~Theisen,
  ``On Scale and Conformal Invariance in Four Dimensions,''
[arXiv:1309.2921 [hep-th]].
%%CITATION = arXiv:1309.2921%%
}

%\IntriligatorIF
\lref\IntriligatorIF{
  K.~A.~Intriligator,
  ``IR free or interacting? A Proposed diagnostic,''
Nucl.\ Phys.\  {\bf B730}, 239-251 (2005).
[hep-th/0509085].
%%CITATION = hep-th/0509085%%
}

%\VartanovXJ
\lref\VartanovXJ{
  G.~S.~Vartanov,
  ``On the ISS model of dynamical SUSY breaking,''
Phys.\ Lett.\  {\bf B696}, 288-290 (2011).
[arXiv:1009.2153 [hep-th]].
%%CITATION = arXiv:1009.2153%%
}

%\GerchkovitzZRA
\lref\GerchkovitzZRA{
  E.~Gerchkovitz,
  ``Constraints on the R-charges of Free Bound States from the R\"omelsberger Index,''
[arXiv:1311.0487 [hep-th]].
%%CITATION = arXiv:1311.0487%%
}

%\MyersXS
\lref\MyersXS{
  R.~C.~Myers and A.~Sinha,
  ``Seeing a c-theorem with holography,''
Phys.\ Rev.\ D {\bf 82}, 046006 (2010).
[arXiv:1006.1263 [hep-th]].
%%CITATION = arXiv:1006.1263%%
}

%\MyersTJ
\lref\MyersTJ{
  R.~C.~Myers, A.~Sinha,
  ``Holographic c-theorems in arbitrary dimensions,''
JHEP {\bf 1101}, 125 (2011).
[arXiv:1011.5819 [hep-th]].
%%CITATION = arXiv:1011.5819%%
}

%\TachikawaTT
\lref\TachikawaTT{
  Y.~Tachikawa and B.~Wecht,
  ``Explanation of the Central Charge Ratio 27/32 in Four-Dimensional Renormalization Group Flows between Superconformal Theories,''
Phys.\ Rev.\ Lett.\  {\bf 103}, 061601 (2009).
[arXiv:0906.0965 [hep-th]].
%%CITATION = arXiv:0906.0965%%
}

%\GaiottoWE
\lref\GaiottoWE{
  D.~Gaiotto,
  ``N=2 dualities,''
JHEP {\bf 1208}, 034 (2012).
[arXiv:0904.2715 [hep-th]].
%%CITATION = arXiv:0904.2715%%
}

%\GreenDA
\lref\GreenDA{
  D.~Green, Z.~Komargodski, N.~Seiberg, Y.~Tachikawa and B.~Wecht,
  ``Exactly Marginal Deformations and Global Symmetries,''
JHEP {\bf 1006}, 106 (2010).
[arXiv:1005.3546 [hep-th]].
%%CITATION = arXiv:1005.3546%%
}

%\MinahanFG
\lref\MinahanFG{
  J.~A.~Minahan and D.~Nemeschansky,
  ``An N=2 superconformal fixed point with E(6) global symmetry,''
Nucl.\ Phys.\ B {\bf 482}, 142 (1996).
[hep-th/9608047].
%%CITATION = hep-th/9608047%%
}

%\GaddeUV
\lref\GaddeUV{
  A.~Gadde, L.~Rastelli, S.~S.~Razamat and W.~Yan,
  ``Gauge Theories and Macdonald Polynomials,''
Commun.\ Math.\ Phys.\  {\bf 319}, 147 (2013).
[arXiv:1110.3740 [hep-th]].
%%CITATION = arXiv:1110.3740%%
}

%\DumitrescuIU
\lref\DumitrescuIU{
  T.~T.~Dumitrescu, N.~Seiberg,
  ``Supercurrents and Brane Currents in Diverse Dimensions,''
JHEP {\bf 1107}, 095 (2011).
[arXiv:1106.0031 [hep-th]].
%%CITATION = PUPT-2372%%
}

%\AlbaYDA
\lref\AlbaYDA{
  V.~Alba and K.~Diab,
  ``Constraining conformal field theories with a higher spin symmetry in d=4,''
[arXiv:1307.8092 [hep-th]].
%%CITATION = arXiv:1307.8092%%
}

%\XieJC
\lref\XieJC{
  D.~Xie and P.~Zhao,
  ``Central charges and RG flow of strongly-coupled N=2 theory,''
JHEP {\bf 1303}, 006 (2013).
[arXiv:1301.0210 [hep-th]].
%%CITATION = DAMTP-2013-1%%
}

%\ArgyresXN
\lref\ArgyresXN{
  P.~C.~Argyres, M.~R.~Plesser, N.~Seiberg and E.~Witten,
  ``New N=2 superconformal field theories in four-dimensions,''
Nucl.\ Phys.\ B {\bf 461}, 71 (1996).
[hep-th/9511154].
%%CITATION = hep-th/9511154%%
}

%\NakayamaIS
\lref\NakayamaIS{
  Y.~Nakayama,
  ``A lecture note on scale invariance vs conformal invariance,''
[arXiv:1302.0884 [hep-th]].
%%CITATION = CALT-68-2910%%
}

%\GiacomelliTIA
\lref\GiacomelliTIA{
  S.~Giacomelli,
  ``Confinement and duality in supersymmetric gauge theories,''
[arXiv:1309.5299 [hep-th]].
%%CITATION = arXiv:1309.5299%%
}

%\XieHS
\lref\XieHS{
  D.~Xie,
  ``General Argyres-Douglas Theory,''
JHEP {\bf 1301}, 100 (2013).
[arXiv:1204.2270 [hep-th]].
%%CITATION = arXiv:1204.2270%%
}

%\KomargodskiPC
\lref\KomargodskiPC{
  Z.~Komargodski, N.~Seiberg,
  ``Comments on the Fayet-Iliopoulos Term in Field Theory and Supergravity,''
JHEP {\bf 0906}, 007 (2009).
[arXiv:0904.1159 [hep-th]].
%%CITATION = arXiv:0904.1159%%
}

%\DobrevQV
\lref\DobrevQV{
  V.~K.~Dobrev and V.~B.~Petkova,
  ``All Positive Energy Unitary Irreducible Representations of Extended Conformal Supersymmetry,''
Phys.\ Lett.\ B {\bf 162}, 127 (1985).
}

%\PapadodimasEU
\lref\PapadodimasEU{
  K.~Papadodimas,
  ``Topological Anti-Topological Fusion in Four-Dimensional Superconformal Field Theories,''
JHEP {\bf 1008}, 118 (2010).
[arXiv:0910.4963 [hep-th]].
%%CITATION = arXiv:0910.4963%%
}

%\PolandEY
\lref\PolandEY{
  D.~Poland, D.~Simmons-Duffin and A.~Vichi,
  ``Carving Out the Space of 4D CFTs,''
JHEP {\bf 1205}, 110 (2012).
[arXiv:1109.5176 [hep-th]].
%%CITATION = arXiv:1109.5176%%
}

%\GreenNQA
\lref\GreenNQA{
  D.~Green and D.~Shih,
  ``Bounds on SCFTs from Conformal Perturbation Theory,''
JHEP {\bf 1209}, 026 (2012).
[arXiv:1203.5129 [hep-th]].
%%CITATION = arXiv:1203.5129%%
}

%\PoppitzKZ
\lref\PoppitzKZ{
  E.~Poppitz, M.~Unsal,
  ``Chiral gauge dynamics and dynamical supersymmetry breaking,''
JHEP {\bf 0907}, 060 (2009).
[arXiv:0905.0634 [hep-th]].
%%CITATION = arXiv:0905.0634%%
}

%\MinwallaKA
\lref\MinwallaKA{
  S.~Minwalla,
  ``Restrictions imposed by superconformal invariance on quantum field theories,''
Adv.\ Theor.\ Math.\ Phys.\  {\bf 2}, 781 (1998).
[hep-th/9712074].
%%CITATION = hep-th/9712074%%
}

%\GaiottoJF
\lref\GaiottoJF{
  D.~Gaiotto, N.~Seiberg and Y.~Tachikawa,
  ``Comments on scaling limits of 4d N=2 theories,''
JHEP {\bf 1101}, 078 (2011).
[arXiv:1011.4568 [hep-th]].
%%CITATION = arXiv:1011.4568%%
}

%\LutyWW
\lref\LutyWW{
  M.~A.~Luty, J.~Polchinski and R.~Rattazzi,
  ``The $a$-theorem and the Asymptotics of 4D Quantum Field Theory,''
JHEP {\bf 1301}, 152 (2013).
[arXiv:1204.5221 [hep-th]].
%%CITATION = arXiv:1204.5221%%
}

%\BeemQXA
\lref\BeemQXA{
  C.~Beem, L.~Rastelli and B.~C.~van Rees,
  ``The N=4 Superconformal Bootstrap,''
Phys.\ Rev.\ Lett.\  {\bf 111}, 071601 (2013).
[arXiv:1304.1803 [hep-th]].
%%CITATION = YITP-SB-13-10%%
}

%\YonekuraKB
\lref\YonekuraKB{
  K.~Yonekura,
  ``Perturbative c-theorem in d-dimensions,''
JHEP {\bf 1304}, 011 (2013).
[arXiv:1212.3028].
%%CITATION = arXiv:1212.3028%%
}

%\FortinHC
\lref\FortinHC{
  J.~-F.~Fortin, B.~Grinstein, C.~W.~Murphy and A.~Stergiou,
  ``On Limit Cycles in Supersymmetric Theories,''
Phys.\ Lett.\ B {\bf 719}, 170 (2013).
[arXiv:1210.2718 [hep-th]].
%%CITATION = arXiv:1210.2718%%
}

%\ShapereZF
\lref\ShapereZF{
  A.~D.~Shapere and Y.~Tachikawa,
  ``Central charges of N=2 superconformal field theories in four dimensions,''
JHEP {\bf 0809}, 109 (2008).
[arXiv:0804.1957 [hep-th]].
%%CITATION = arXiv:0804.1957%%
}

%\ElvangST
\lref\ElvangST{
  H.~Elvang, D.~Z.~Freedman, L.~-Y.~Hung, M.~Kiermaier, R.~C.~Myers and S.~Theisen,
  ``On renormalization group flows and the a-theorem in 6d,''
JHEP {\bf 1210}, 011 (2012).
[arXiv:1205.3994 [hep-th]].
%%CITATION = arXiv:1205.3994%%
}

%\ZhiboedovOPA
\lref\ZhiboedovOPA{
  A.~Zhiboedov,
  ``On Conformal Field Theories With Extremal a/c Values,''
[arXiv:1304.6075 [hep-th]].
%%CITATION = arXiv:1304.6075%%
}

%\FortinHN
\lref\FortinHN{
  J.~-F.~Fortin, B.~Grinstein and A.~Stergiou,
  ``Limit Cycles and Conformal Invariance,''
JHEP {\bf 1301}, 184 (2013), [JHEP {\bf 1301}, 184 (2013)].
[arXiv:1208.3674 [hep-th]].
%%CITATION = arXiv:1208.3674%%
}

%\DolanZH
\lref\DolanZH{
  F.~A.~Dolan and H.~Osborn,
  ``On short and semi-short representations for four-dimensional superconformal symmetry,''
Annals Phys.\  {\bf 307}, 41 (2003).
[hep-th/0209056].
%%CITATION = hep-th/0209056%%
}

%\NakayamaND
\lref\NakayamaND{
  Y.~Nakayama,
  ``Supercurrent, Supervirial and Superimprovement,''
Phys.\ Rev.\ D {\bf 87}, 085005 (2013).
[arXiv:1208.4726 [hep-th]].
%%CITATION = IPMU12-0160%%
}

%\NakayamaWX
\lref\NakayamaWX{
  Y.~Nakayama,
  ``Higher derivative corrections in holographic Zamolodchikov-Polchinski theorem,''
[arXiv:1009.0491 [hep-th]].
%%CITATION = arXiv:1009.0491%%
}

%\KomargodskiXV
\lref\KomargodskiXV{
  Z.~Komargodski,
  ``The Constraints of Conformal Symmetry on RG Flows,''
JHEP {\bf 1207}, 069 (2012).
[arXiv:1112.4538 [hep-th]].
%%CITATION = arXiv:1112.4538%%
}

%\LiuEEA
\lref\LiuEEA{
  H.~Liu and M.~Mezei,
  ``A Refinement of entanglement entropy and the number of degrees of freedom,''
[arXiv:1202.2070 [hep-th]].
%%CITATION = MIT-CTP-4336%%
}

%\JafferisZI
\lref\JafferisZI{
  D.~L.~Jafferis, I.~R.~Klebanov, S.~S.~Pufu and B.~R.~Safdi,
  ``Towards the F-Theorem: N=2 Field Theories on the Three-Sphere,''
JHEP {\bf 1106}, 102 (2011).
[arXiv:1103.1181 [hep-th]].
%%CITATION = arXiv:1103.1181%%
}

%\KutasovXB
\lref\KutasovXB{
  D.~Kutasov,
  ``Geometry on the Space of Conformal Field Theories and Contact Terms,''
Phys.\ Lett.\ B {\bf 220}, 153 (1989).
%%CITATION = WIS-88-55-PH%%
}

%\SeibergUR
\lref\SeibergUR{
  N.~Seiberg,
  ``Supersymmetry and Nonperturbative beta Functions,''
Phys.\ Lett.\ B {\bf 206}, 75 (1988).
%%CITATION = IASSNS-HEP-88-5%%
}

%\ErkalSH
\lref\ErkalSH{
  D.~Erkal and D.~Kutasov,
  ``a-Maximization, Global Symmetries and RG Flows,''
[arXiv:1007.2176 [hep-th]].
%%CITATION = arXiv:1007.2176%%
}

%\FarnsworthOSA
\lref\FarnsworthOSA{
  K.~Farnsworth, M.~A.~Luty and V.~Prelipina,
  ``Scale Invariance plus Unitarity Implies Conformal Invariance in Four Dimensions,''
[arXiv:1309.4095 [hep-th]].
%%CITATION = arXiv:1309.4095%%
}

%\CasiniEI
\lref\CasiniEI{
  H.~Casini and M.~Huerta,
  ``On the RG running of the entanglement entropy of a circle,''
Phys.\ Rev.\ D {\bf 85}, 125016 (2012).
[arXiv:1202.5650 [hep-th]].
%%CITATION = arXiv:1202.5650%%
}

%\AharonyDJ
\lref\AharonyDJ{
  O.~Aharony and Y.~Tachikawa,
  ``A Holographic computation of the central charges of d=4, N=2 SCFTs,''
JHEP {\bf 0801}, 037 (2008).
[arXiv:0711.4532 [hep-th]].
%%CITATION = arXiv:0711.4532%%
}

%\CaraccioloBX
\lref\CaraccioloBX{
  F.~Caracciolo and V.~S.~Rychkov,
  ``Rigorous Limits on the Interaction Strength in Quantum Field Theory,''
Phys.\ Rev.\ D {\bf 81}, 085037 (2010).
[arXiv:0912.2726 [hep-th]].
%%CITATION = arXiv:0912.2726%%
}

%\HookFP
\lref\HookFP{
  A.~Hook,
  ``A Test for emergent dynamics,''
JHEP {\bf 1207}, 040 (2012).
[arXiv:1204.4466 [hep-th]].
%%CITATION = arXiv:1204.4466%%
}

%\BertoliniVKA
\lref\BertoliniVKA{
  M.~Bertolini, L.~Di Pietro and F.~Porri,
  ``Holographic R-symmetric flows and the $\tau_{U}$ conjecture,''
JHEP {\bf 1308}, 071 (2013).
[arXiv:1304.1481 [hep-th]].
%%CITATION = arXiv:1304.1481%%
}

%\DouglasIC
\lref\DouglasIC{
  M.~R.~Douglas,
  ``Spaces of Quantum Field Theories,''
[arXiv:1005.2779 [hep-th]].
%%CITATION = arXiv:1005.2779%%
}

\rightline{RU-NHETC-2013-21}
\Title{\vbox{\baselineskip12pt }} {\vbox{\centerline{Minimal Distances Between SCFTs}}}
\smallskip
\centerline{Matthew Buican}
\smallskip
\bigskip
\centerline{{\it Department of Physics and Astronomy, Rutgers University, Piscataway, NJ 08854, USA}}
\vskip 1cm
\noindent
We study lower bounds on the minimal distance in theory space between four-dimensional superconformal field theories (SCFTs) connected via broad classes of renormalization group (RG) flows preserving various amounts of supersymmetry (SUSY). For $\CN=1$ RG flows, the ultraviolet (UV) and infrared (IR) endpoints of the flow can be parametrically close. On the other hand, for RG flows emanating from a maximally supersymmetric SCFT, the distance to the IR theory cannot be arbitrarily small regardless of the amount of (non-trivial) SUSY preserved along the flow. The case of RG flows from $\CN=2$ UV SCFTs is more subtle. We argue that for RG flows preserving the full $\CN=2$ SUSY, there are various obstructions to finding examples with parametrically close UV and IR endpoints. Under reasonable assumptions, these obstructions include: unitarity, known bounds on the $c$ central charge derived from associativity of the operator product expansion, and the central charge bounds of Hofman and Maldacena. On the other hand, for RG flows that break $\CN=2\to\CN=1$, it is possible to find IR fixed points that are parametrically close to the UV ones. In this case, we argue that if the UV SCFT possesses a single stress tensor, then such RG flows excite of order all the degrees of freedom of the UV theory. Furthermore, if the UV theory has some flavor symmetry, we argue that the UV central charges should not be too large relative to certain parameters in the theory.

\bigskip
\Date{November 2013}

\newsec{Introduction}
Quantum field theory (QFT) describes an astonishing range of physics: from spin systems at criticality to colliding particles to quantum gravity in Anti de Sitter space. Given this remarkable diversity and reach, one particularly interesting (and difficult) problem is to describe the space of possible QFTs, commonly referred to as \lq\lq theory space." A slightly simpler though still daunting problem is to characterize the space of conformal field theories (CFTs)---see \DouglasIC\ for a recent discussion.\foot{Throughout this work we will specialize to four dimensions, but some of the statements we make have analogs in other dimensions.}

Individual CFTs can be described in terms of a relatively simple set of numerical data.
A particularly important subset of the CFT data describes the aggregate behavior of the theory: for example, the conformal anomaly coefficients measure the CFT's collective response to gravity (in a curved four-dimensional background, the conformal anomaly contains the Euler density and the square of the Weyl tensor with coefficients $a$ and $c$ respectively), while quantities like the two-point function coefficients of the internal symmetry currents, $\tau_{ij}$, measure the collective response of the CFT to the presence of a weakly coupled gauge field.

The CFT data also obeys various simple physical consistency conditions. Of these many constraints, some important ones we will encounter again below include (note that we will only study unitary theories):

\smallskip
\noindent
$\bullet$ The $c$ anomaly coefficient and the current two-point functions, $\tau_{ij}$, are constrained to be positive definite by unitarity.

\smallskip
\noindent
$\bullet$ From the positivity of energy flux correlators, the ratio of $a$ and $c$ should fall in a certain finite and positive range of values \HofmanAR\ (thus rendering $a$ positive as well).\foot{See \ZhiboedovOPA\ for an interesting recent analysis of theories that saturate these bounds.}

\smallskip
\noindent
$\bullet$ Associativity of the operator product expansion (OPE) applied to the four point function of a real Lorentz scalar primary, $\CO$, of dimension $d$ implies that $c$ (and therefore, by the previous condition, $a$) is bounded from below by some universal positive function of $d$, $f(d)$ \refs{\PolandWG, \RattazziGJ}. Furthermore, the size of the OPE coefficients of operators, $\CO'$, of dimension $D$ appearing in the $\CO\CO$ OPE are bounded from above by some function $M(d, D)$ \CaraccioloBX\ (here we are taking a basis of canonically normalized primaries; note that $f(d)$ and $M(d,D)$ have only been numerically computed for certain ranges of $d$ and $D$, but it seems plausible to extend the analysis to larger ranges of values).

\smallskip
\noindent
Heuristically, the last bullet point amounts to saying that there should be some minimal (non-zero) amount of stuff present in a unitary theory and that the CFT cannot be arbitrarily strongly coupled (at least in the precise sense above).

CFTs are also useful because they describe the endpoints of RG flows.\foot{In this paper we only consider UV-complete quantum field theories (QFTs). This requirement guarantees that the endpoints of the RG flow are scale invariant (although possibly trivial in the IR). Furthermore, under rather mild assumptions, such theories must also be conformal in two dimensions \PolchinskiDY. In four dimensions the situation is more complicated, but recent work has shed a great deal of light on this question \refs{\NakayamaWX\DorigoniRA\AntoniadisGN\LutyWW\FortinHC\NakayamaND\FortinHN\NakayamaIS\DymarskyPQA-\FarnsworthOSA} (the relationship between scale and conformal invariance has also been explored in other dimensions \YonekuraKB), and it is clear that if unitary interacting scale invariant (but non-conformal) theories exist, they must be very special (see, in particular, the recent non-perturbative conditions derived in \DymarskyPQA). Note that the free two-form theory in four dimensions is unitary, scale-invariant, and non-conformal---the authors of \refs{\DymarskyPQA} have emphasized that this conclusion holds for the case of a non-compact gauge symmetry; this theory is also special since it does not have a well-defined scaling current.} We can then compare the constrained numerical CFT data in the deep UV and the deep IR and search for laws that describe the RG flow and the part of theory space that it probes. The seminal result in this line of research is Zamolodchikov's $c$-theorem for two-dimensional RG flows \ZamolodchikovGT, which states that a flow connects two CFTs---$\CT_{UV}$ at short distances with $\CT_{IR}$ at long distances---only if the respective central charges, $c_{UV}$, and, $c_{IR}$, satisfy
\eqn\cthm{
c_{UV}>c_{IR}~.
}
Note that in our definition of the RG flow, we imagine that we introduce a scale into the theory by turning on some relevant deformation, and so we exclude motion along a manifold of fixed points (a so-called \lq\lq conformal manifold")---such motion leaves $c$ invariant. Clearly, the $c$-theorem introduces an ordering in the space of two dimensional CFTs (or, more precisely, into the space of conformal manifolds).

In fact, Zamolodchikov proved slightly more. In particular, he constructed a $c$ function that decreases monotonically along the RG flow (depending on the RG scale only through the couplings of the theory) and interpolates between $c_{UV}$ and $c_{IR}$ (this $c$-function is stationary in the deep UV and the deep IR).

One can naturally conjecture a generalization of \cthm\ to RG flows in higher dimensions. In $d$ even dimensions, the natural generalization of $c$ is the coefficient of the logarithmically divergent piece of the partition function of the theory on the $d$-sphere, $S_d$, while in $d$ odd dimensions, the natural generalization is the constant piece of the partition function on $S_d$ (this term is referred to in the literature as \lq\lq$F$," and it has been argued to obey an inequality---conjectured by various authors \refs{\MyersXS\JafferisZI\MyersTJ-\LiuEEA}---analogous to \cthm\ in three dimensions \CasiniEI).

Since we will be working in four dimensions throughout this paper, the most relevant generalization of \cthm\ for our purposes is the $a$-theorem (originally conjectured in \CardyCWA) of Komargodski and Schwimmer \refs{\KomargodskiVJ, \KomargodskiXV} which states that the four-dimensional RG flow connects a UV theory, $\CT_{UV}$, with an IR theory, $\CT_{IR}$, only if the $a$ central charge we introduced above (note that $a$ can also be defined as the coefficient of the logarithmically divergent part of the partition function of the theory on $S_4$), satisfies
\eqn\athm{
a_{UV}>a_{IR}~.
}
As in the case of Zamolodchikov's $c$-theorem in two dimensions, the $a$-theorem implies an ordering on the space of four-dimensional CFTs.\foot{Generalizing the $c$-theorem to other dimensions remains an open problem (see, e.g. \ElvangST).} Note, however, that there is not as yet a direct generalization of Zamolodchikov's $c$-function to a monotonically decreasing $a$-function depending on the RG scale only through the couplings (it may well be the case that such a function can be constructed using elements introduced in \KomargodskiVJ).

Another useful aspect of CFTs is that they define natural metrics in the patches of theory space to which they belong (see also \KutasovXB). These metrics are matrices of two-point functions for primaries in the CFT, 
\eqn\zamomet{
G^0_{IJ}\sim\langle\CO_I(x)\CO_J(0)\rangle\cdot x^{2D_I}~,
}
where $D_I$ is the scaling dimension of $\CO_I$ (in writing the above metric, we are imagining that the $\CO_I$ are hermitian; therefore, unitarity constrains this metric to be positive definite). Slightly away from criticality, we can use conformal perturbation theory to find the radiative corrections to $G^0_{IJ}$. Extending \zamomet\ globally away from a given CFT is a very subtle problem, and there may be other metrics that are more useful for this purpose (see the discussion in \DouglasIC), but these global issues will be immaterial for us below and are beyond the scope of this work. We simply note that \zamomet\ provides a reasonable notion of distance in theory space in the vicinity of a given CFT, and we will assume that we can find some appropriate generalization of distance in the rest of theory space that locally approximates distances computed with \zamomet.

The above discussion is rather general and does not assume the existence of additional symmetries beyond the conformal group (at very special points in the space of theories). It is therefore reasonable to expect that we should be able to derive more powerful results if we study certain subclasses of theories with more symmetry. For example, in \refs{\BuicanTY} we conjectured and tested an additional constraint on four-dimensional RG flows that are supersymmetric (and also $R$-symmetric).\foot{See \BertoliniVKA\ for an interesting holographic interpretation. In \BuicanEC\ we studied extensions of \BuicanTY\ to theories with small amounts of SUSY breaking.}  This constraint gives rise to predictions about the IR phases of certain supersymmetric theories that are not constrained by the $a$-theorem.\foot{See \refs{\IntriligatorIF\HookFP\VartanovXJ\PoppitzKZ-\GerchkovitzZRA} for additional proposed constraints on supersymmetric RG flows.}
\bigskip
\centerline{
\includegraphics[height=1.75in,width=2.25in,angle=0]{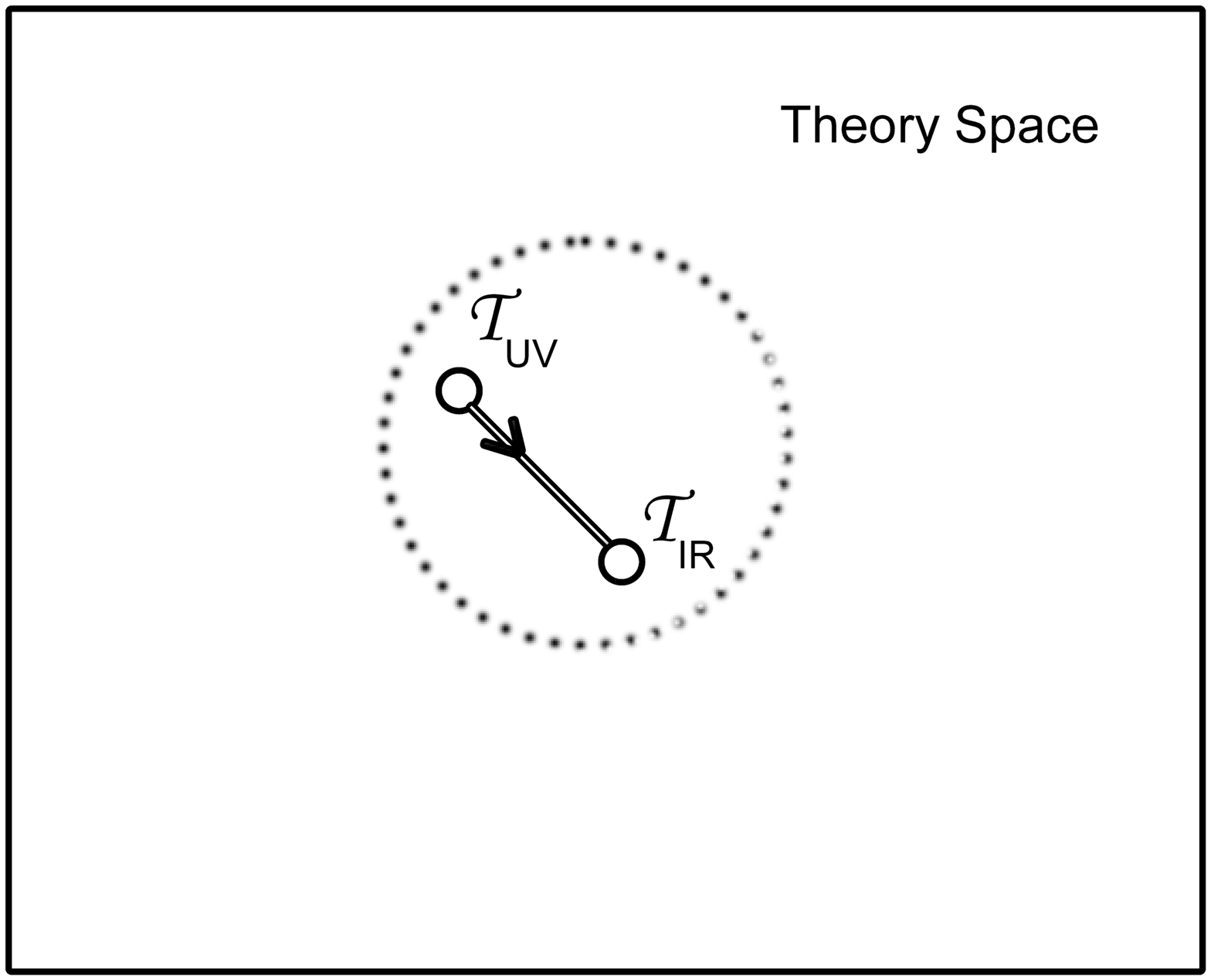}
\includegraphics[height=1.75in,width=2.25in,angle=0]{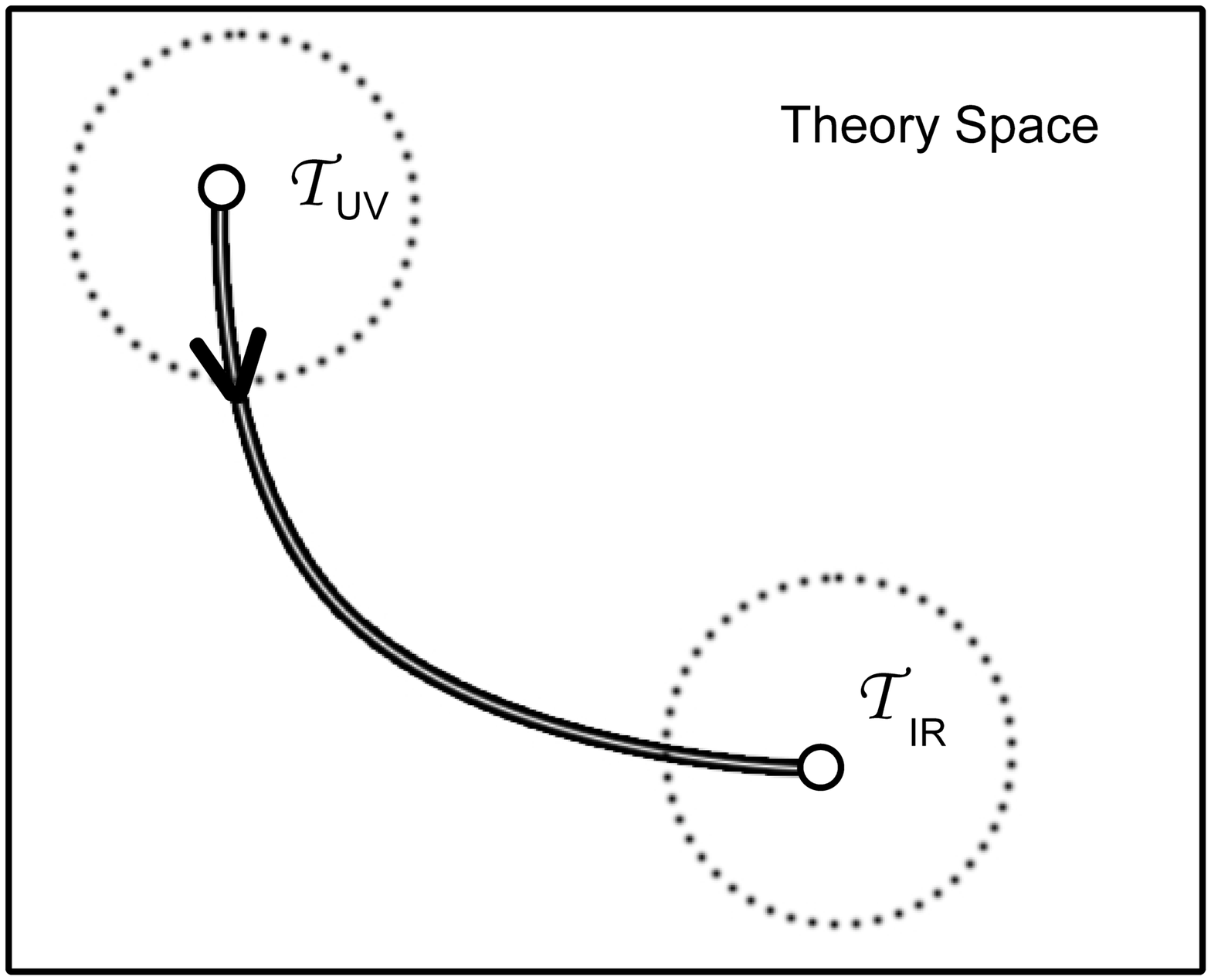}}
\centerline{
{\bf Figure 1:} For theories with less supersymmetry, we expect that the dynamics\ }\centerline{in the vicinity of the UV theory is richer and that the RG flow can often conn-}\centerline{ect multiple fixed points in the same neighborhood (left). On the other hand, \ }\centerline{in theories with more supersymmetry, we expect that the flow typically comes }\centerline{to a stop due to dynamics in another patch of theory space (right). Note that }\centerline{$\CT_{UV,IR}$ may be part of some (distinct) non-trivial conformal manifolds. \ \ \ \ \ \ \ \ \ }
\bigskip

In this note, we are interested in understanding how \lq\lq repulsive" SCFTs are in the space of supersymmetric theories. More concretely, we start with some UV SCFT, $\CT_{UV}$, and we turn on a SUSY-preserving relevant deformation that initiates an RG flow to a new SUSY fixed point, $\CT_{IR}$ (with the brief exception of studying motion on the moduli space of the $\CN=4$ theory, we will mostly avoid discussing RG flows that consist of breaking conformal symmetry spontaneously in the UV CFT; we will typically treat any vevs for CFT operators as small perturbations of our relevant deformations in the UV). We then search for lower bounds on the distance between the end points of the RG flow in theories with differing amounts of SUSY (see figure 1). Intuitively, we expect that the more supersymmetric our theory is, the more constrained the short-distance dynamics and therefore the more likely it is that the resulting RG flows will be forced to traverse a large distance in theory space (we expect that some non-perturbative dynamics in a different patch of theory space will usually be needed to bring the RG flow to a stop). In making these notions more precise, we will make contact with many general properties of CFTs, including various theory-independent bounds on the central charges of CFTs (some of which we have already mentioned above).

Note that we can define the \lq\lq distance" between $\CT_{UV}$ and $\CT_{IR}$ in many ways. As we discussed above, one way that makes sense in the vicinity of $\CT_{UV}$ is to use the metric in \zamomet. In particular, if the IR theory can be described in terms of UV operators with some small changes in couplings, $\delta\lambda^I$, then we can characterize the distance between the theories as
\eqn\dist{
d(\CT_{UV},\CT_{IR})=\int_{\gamma}\left(G_{IJ}d\lambda^Id\lambda^J\right)^{1\over2}\sim\left(G^0_{IJ}\delta\lambda^I\delta\lambda^J\right)^{1\over2}=\left(\delta_{IJ}\delta\lambda^I\delta\lambda^J\right)^{1\over2}=d_0(\CT_{UV},\CT_{IR})~,
}
where we have chosen an orthonormal basis at $\CT_{UV}$, i.e.  $G^0_{IJ}=\delta_{IJ}$. In \dist, $\gamma$ is a minimal length geodesic connecting $\CT_{UV}$ and $\CT_{IR}$ in the space of theories, and $d_0$ is an approximation for very small changes in the $\lambda^I$.\foot{Note that $d(\CT_{UV}, \CT_{IR})$ is scheme independent since it is diffeomorphism invariant in the space of couplings. Of course, we should interpret \dist\ with care, since theory space is infinite dimensional (one simplification we can make is to ignore the highly irrelevant directions around $\CT_{UV}$).} Presumably \dist\ is well-defined if $\CT_{IR}$ is within some small but finite universal distance $D_{\rm univ}>d(\CT_{UV}, \CT_{IR})$ from $\CT_{UV}$.\foot{Note that we are assuming $G^0_{IJ}$ is well-defined; therefore, we will not discuss theories with free gauge fields.} 
Otherwise, we would find that there are theories with arbitrarily small changes in couplings of finitely normalized operators where \dist\ is no longer well-defined, and the IR theory should be written in terms of some new degrees of freedom, i.e., $\CT_{IR}$ is in a different patch of theory space. Such a situation would occur if we are forced to consider RG flows between $\CT_{UV}$ and $\CT_{IR}$ with arbitrarily large beta functions in the vicinity of $\CT_{UV}$ (measured in a scheme-independent way via the metric) or if $\CT_{UV}$ has arbitrarily large OPE coefficients for unit normalized operators. Recall from our above discussion that associativity of the OPE applied to scalar four-point functions already requires certain upper bounds on OPE coefficients \refs{\CaraccioloBX, \PolandEY}. This fact suggests that the physics in neighborhoods of CFTs should be smooth.

Another potentially interesting measure of the distance between the endpoints of the RG flow is the change in the $a$ anomaly, $\delta a=a_{UV}-a_{IR}$ (see, e.g. \IntriligatorIF).\foot{In \AmaritiWC, the authors discussed RG flows with $\delta a=0$ and so we should interpret $\delta a$ as a measure of the length of the RG flow with care. For our purposes here, we simply note that the $\delta a=0$ RG flows of \AmaritiWC\ are special in the sense that they can be understood as being initiated by giving vevs to scalars that only interact with the UV SCFT via highly irrelevant operators. In particular, turning on these vevs does not generate relevant deformations of the UV theory.} This measure of distance has the advantage that it can often be computed exactly in SUSY RG flows (sometimes even if $\CT_{IR}$ is in a different patch of theory space, as in the case of SQCD in the free magnetic phase), and we can then test whether $\delta a$ is bounded positively from below or not in different classes of theories.\foot{Note that $\delta a$ is a measure of distance only in directions of theory space normal to conformal manifolds (since we have $\delta a=0$ for exactly marginal deformations). On the other hand, if we want to consider motion along a conformal manifold, then we can still use \dist\ as a measure of distance. However, in this paper, we are only interested in motion normal to conformal manifolds.}

In fact, it is reasonable to assume that bounds on $\delta a$ and bounds on $d$ are related (note that if the RG flow probes a new patch of theory space and \dist\ is not well-defined, then we simply say that $d\ge D_{\rm univ}$, where we have defined $D_{\rm univ}$ below \dist).  For example, if we can establish that $\delta a>\Delta_{\rm univ}$ for a class of RG flows (here $\Delta_{\rm univ}>0$ is a universal constant for the class), then we expect that $d>d_{\rm univ}$ for some universal $0<d_{\rm univ}<D_{\rm univ}$. Indeed, if this were not the case, then we would find theories in the class with $d$ arbitrarily small and $\delta a$ finite. Such a situation would be incompatible with continuity (note that in the neighborhood of the CFT, we can write an $a$-function that is closely related to \dist\foot{$a$-functions in the vicinity of CFTs have been much-discussed in the recent literature, see \NakayamaIS\ for a review.}).

As we will see in the subsequent sections, minimally supersymmetric RG flows can connect parametrically close UV and IR SCFTs (in both senses we described above), while RG flows that preserve some non-minimal SUSY cannot. More precisely, we will show that---under reasonable assumptions---for RG flows with $\CN=2$ SUSY at least one of the following applies

\smallskip
\noindent
{\bf(i)} $\delta a>\Delta_{\rm univ}>0$.

\smallskip
\noindent
{\bf(ii)} The IR superconformal $R$ current must mix with currents that are badly broken in the UV SCFT, i.e., currents that have short-distance dimension $3+\gamma$ with $\gamma>\gamma_{\rm univ}>0$ (here $\gamma_{\rm univ}$ is a universal positive number).

\smallskip
\noindent
{\bf(iii)} $\CT_{IR}$ is in a different patch of theory space than $\CT_{UV}$. In particular, the short-distance and long-distance degrees of freedom are necessarily different.

\smallskip
\noindent
{\bf (iv)} The RG flow proceeds along directions with an ill-defined Zamolodchikov metric.

\smallskip
\noindent
This result is already enough to show that $d>d_{\rm univ}$, since otherwise we would again run into the requirement that there are RG flows with arbitrarily large UV beta functions (measured via $G_{IJ}$) or UV SCFTs with arbitrarily large OPE coefficients for unit normalized operators.\foot{One interesting example of {\bf(iii)} occurs when some unbroken non-abelian symmetries of the UV theory decouple in the IR. Indeed, if one can generalize the discussion in \PolandEY\ on bounds for non-abelian current two-point functions (more precisely, these bounds are on ratios of the two-point function coefficients to the indices of the representations of certain charged operators) to theories with arbitrary non-abelian symmetry groups and matter representations, then the UV non-abelian two-point function coefficients, $\tau_{ij}^{UV}>0$, must be bounded positively (and universally) from below (more precisely, this bound would be on the ratio of the two-point function coefficients to the indices of the representations for low-lying scalar operators in the theory). In this case, we expect that $\tau_{ij}^{UV}$ cannot flow to zero for arbitrarily small $d$.} Furthermore, if we assume that there exists some $a$-function that interpolates between $a_{UV}$ and $a_{IR}$ and is not \lq\lq fine-tuned," then we can conclude that $\delta a$ is always bounded positively from below in RG flows with non-minimal SUSY.

If we start with an $\CN=2$ supersymmetric theory and explicitly break it to $\CN=1$, then it is possible that the RG flow can be parametrically short (note that we could also imagine studying RG flows that are $\CN=1$ supersymmetric except for an accidental $\CN=2$ SUSY appearing in the deep IR; we will not discuss such RG flows in this work). However, if the $\CN=2$ theory has some flavor symmetry and a single stress tensor (i.e., a single sector), then there are constraints on how large the central charges can be relative to the inverse of the deviation from marginality of the relevant deformations that initiate the flow. In these cases (modulo an assumption we discuss below), the RG flow necessarily excites of order all the degrees of freedom that contribute to the central charges of the UV theory.

Let us briefly discuss the plan for the rest of the paper. In the next section, we describe some of the general aspects of the RG flows we study. In section three we discuss RG flows with minimal SUSY, while in section four we discuss RG flows emanating from $\CN=2$ SCFTs (we consider RG flows that preserve both the full $\CN=2$ SUSY as well as those that preserve only an $\CN=1$ subalgebra). In section five we briefly discuss RG flows from $\CN=4$ theories, and then we conclude in section six. In the appendix we prove a simple and useful result on the absence of mixing of almost conserved symmetries of $\CN=2$ UV SCFTs with the IR superconformal $R$ current under certain assumptions that we describe in detail.

\newsec{General aspects of the SUSY RG flow}
In this section we would like to describe some general aspects of the SUSY RG flows we study. We start by taking some UV SCFT, $\CT_{UV}$, and turning on a relevant deformation (we assume that any non-trivial vevs for CFT operators are sub-leading deformations at short distances). Unitarity constrains this deformation to be of the form\foot{Note that asymptotically free gauge fields do not lead to parametrically small $d$ (they do not lead to parametrically small $\delta a$ either if we assume that there is a non-fine-tuned decreasing $a$-function). Indeed, in these cases, the Zamolodchikov metric is singular (see, e.g., \KomargodskiXV). Such deformations are therefore not of interest to us. We will also not consider deforming by operators that are, on their own, (marginally) irrelevant but become relevant when we add a genuinely relevant operator to the action. Much of what we will say carries over to this case, but there are a few caveats.}
\eqn\reldef{
\delta W=\lambda^i\mu^{3-D_i}\CO_i~, \ \ \ D_i=3-\Delta_i~,
}
where $\CO_i$ is a relevant chiral primary of dimension $D_i$ (here $0<\Delta_i\le2$ is the deviation from marginality). Note that even though we will discuss theories with non-minimal SUSY, it is useful to use $\CN=1$ superspace notation to describe our theories. For simplicity, we will assume that our chiral primaries are canonically normalized
\eqn\canonnorm{
g_{i\bar j}=\langle\CO_i(x)\CO^{\dagger}_{\bar j}(0)\rangle\cdot x^{2D_i}=\delta_{i\bar j}~.
}

In the vicinity of the UV theory, the physics is controlled by the operator product expansions (OPEs) for the deformations
\eqn\OPEs{
\CO_i(x)\CO_{\bar j}^{\dagger}(0)={\delta_{i\bar j}\over|x|^{6-2\Delta_i}}+{T^A_{i\bar j}\over|x|^{4-\Delta_i-\Delta_j}} J_A+{c^I_{i\bar j}\over|x|^{6-D_I-\Delta_i-\Delta_j}}L_I+\cdot\cdot\cdot~.
}
Here the $L_I$ are scalar primaries of dimension $D_I>2$, and the $J_A$ (satisfying $\left[Q^2,J_A\right]=\left[\bar Q^2, J_A\right]=0$) are dimension two primary operators corresponding to the flavor (from the $\CN=1$ perspective, this means non-$U(1)_R$) symmetries of $\CT_{UV}$ that the relevant deformations are charged under (the $J_A$ are sometimes called \lq\lq moment maps" for the corresponding symmetries). The $J_A$ operators are related by SUSY to conserved symmetry currents.\foot{The ellipses in \OPEs\ include higher-dimensional descendants of the $L_I$ and the $J_A$---in particular, the corresponding spin one conserved currents, $j_{A\mu}$, which sit in the $\theta\bar\theta$ components of the $J_A$ superfields and various non-conserved vector operators $L_{I\mu}$. Note that the $L_{I\mu}$ may include approximately conserved currents of the UV theory. We will discuss such operators further below.} The $T^A_{i\bar j}$ specify the corresponding charges. More precisely,
\eqn\Tibarjdef{
T^A_{i\bar j}=4\pi^2\cdot\tau^{AB}\left(t_B\right)_i^k\delta_{k\bar j}~,
}
where $\tau^{AB}$ is the inverse of the Zamolodchikov metric for the moment maps, and $\left(t_A\right)_i^k$ is a representation matrix for the symmetries corresponding to $J_A$ (with charge $Q_A$) acting on $\CO_k$
\eqn\Zamomom{
\tau_{AB}=\langle J_A(x)J_B(0)\rangle\cdot(2\pi)^4\cdot|x|^4>0~, \ \ \ \left[Q_A, \CO_i\right]=-\left(t_A\right)_i^j\CO_j~.
}
As we go to long distances, we expect that the description of the physics in terms of the OPEs in \OPEs\ will typically break down unless the IR theory is in the same patch of theory space as the UV SCFT.

From the general discussion in the introduction, it is clear that in order to search for RG flows with parametrically small distance between the UV and IR endpoints as measured by $d\sim d_0$ in \dist, we should study RG flows initiated by almost marginal relevant deformations, i.e., deformations with $\Delta_i\ll1$ (otherwise we would again need to find UV theories with arbitrarily large OPE coefficients for finitely normalized operators or RG flows with arbitrarily large UV beta functions measured with $g_{i\bar j}$ so that the relevant deformation flows to a marginal or irrelevant operator in the IR).

We can examine such theories in more detail by using the OPEs in \OPEs\ to compute the low-order beta functions (we impose a short-distance cutoff on operator collisions in superspace). For simplicity, let us consider the case of a UV SCFT with a single $U(1)$ flavor symmetry, $U$, and let us turn on a single relevant deformation, $\delta W=\lambda\mu^{\Delta}\CO$, with $\Delta=\epsilon\ll1$ and charge $U(\CO)=\epsilon$. In this case, we find that the physical (as opposed to holomorphic) coupling changes according to
\eqn\betai{
\beta=-\epsilon\lambda+4\pi^4{\epsilon^2\over\tau_U}\lambda^3+\cdot\cdot\cdot~,
}
where we have taken $\lambda$ to be real, and $\tau_U$ is the two-point function of $U$ in the UV SCFT. Note that in this case, the OPE \OPEs\ takes the form $\CO(x)\CO^{\dagger}(0)=\cdot\cdot\cdot+{\epsilon\over\sqrt{\tau_U} |x|^{4-2\epsilon}}\cdot{4\pi^2 U(0)\over\sqrt{\tau_U}}+\cdot\cdot\cdot$, and the OPE coefficient bounds of \CaraccioloBX\ require
\eqn\bounds{
{\epsilon\over\sqrt{\tau_U}}<M(D_{\CO}, 2)~,
}
where $M$ is a universal (finite) function depending on the UV dimension of $\CO$, $D_{\CO}$, and the UV dimension of $U$ (which is two). Alternatively, we may simply note that the upper bounds of \PolandWG\ on the ratio of operator charges squared to current two-point function coefficients preclude the quantity in \bounds\ from being arbitrarily large.\foot{Technically, one needs to extend the analyses of \CaraccioloBX\ and \PolandWG\ to higher $D_{\CO}$, since here $D_{\CO}\sim3$, but such an extension does not seem to be problematic.}  From this discussion, we see that
\eqn\dzerocomp{
d\sim d_0(\CT_{UV}, \CT_{IR})=\lambda_*=\left({\tau_U\over4\pi^4\epsilon}\right)^{1\over2}~.
}
Provided that $\tau_U\sim\epsilon^N$ with $1<N\lesssim 2$, $d_0$ can be parametrically small, and we can trust our perturbative analysis. The upper bound on $N$ follows from \bounds.

From the general discussion in the introduction, this RG flow should then also have parametrically small $\delta a$. Indeed, using the formula for the leading-order change in $a$, we see that
\eqn\confperta{
\delta a\sim-{64\pi^4\over3}\int_{0}^{\lambda_*} d\lambda\cdot\beta={4\over3}\tau_U~,
}
where this equation implicitly includes a factor of the Zamolodchikov metric (here trivially set to unity). Note that we have the hierarchy $\delta a\sim\epsilon^{N+1\over2} \cdot d_0\ll d_0$, which reflects the fact that the entire RG flow occurs in the neighborhood of an SCFT. We expect that for more general RG flows---where most of the RG evolution is not spent in the vicinity of an SCFT---the flow of $a$ is not fine-tuned.\foot{Note that in arriving at the expressions in \dzerocomp\ and \confperta\ we have assumed that the relevant deformation does not back-react on the theory. For example, we have neglected higher-order perturbative corrections or the possibility that the relevant deformation may lead to a non-perturbative renormalization of the superpotential or a shift in the vacuum of the theory (one could even imagine that the relevant deformation triggers dynamics that lead to spontaneous SUSY breaking or a run-away potential). Such corrections would generically push $d$ to larger values.}

Using 't Hooft anomaly matching, we can write an exact expression for $\delta a$ as long as the perturbative and non-perturbative corrections to the RG flow are sufficiently small. More precisely, we note that the $U$ symmetry current superfield defines an RG-conserved $R$ symmetry current via \KomargodskiRB
\eqn\Udef{
\bar D^{\dot\alpha}R_{\alpha\dot\alpha}=\bar D^2D_{\alpha}U~.
}
In components, we have that $R_{\mu}=\tilde R_{\mu}^{UV}+{2\over3}U_{\mu}$, where $R_{\mu}$ is the RG-conserved $R$ current, and $\tilde R_{\mu}^{UV}$ is the UV superconformal $R$ current.\foot{In more general settings, the $U$ operator is defined (modulo unimportant shifts) when the theory has an FZ multiplet \KomargodskiRB\ (most theories have such a multiplet, with the exception of theories with FI terms \KomargodskiPC\ or non-trivial target space topology). See \AbelWV\ for some recent applications of this multiplet and \DumitrescuIU\ for an additional general discussion.} In the IR, we expect that this current multiplet flows to $\tilde R_{\mu}^{IR}$, and, using the fact that $a=3\Tr\tilde R^3-\tilde R$, 't Hooft anomaly matching implies the exact result\foot{Note that our conventions imply that $a$ is normalized such that $a={2\over9}$ for a free chiral superfield.}
\eqn\deltafull{
\delta a={4\over3}\tau_U-{8\over9}\Tr U^3~.
}

If we study an RG flow from a more general $\CT_{UV}$ with a more complicated symmetry group, then we should also take these symmetries into account when computing $a_{IR}$. More precisely, these symmetries introduce an ambiguity in \Udef\ under which $U\to U_t=U+t^aJ_a$ and $R\to R_t=R+{2\over3}t^aJ_a$ (here the $J_a$ are the additional symmetry currents of $\CT_{UV}$ that are preserved by the relevant deformations that take us to $\CT_{IR}$). We can compute the IR superconformal $R$ symmetry by maximizing the trial $a$ anomaly functional \IntriligatorJJ
\eqn\amax{
a_t=3\Tr R_t^3-\Tr R_t~,
}
with respect to the $t^a$. This procedure then defines a corresponding $U$ operator that we will simply denote as $U$, and it is this operator that then appears in the generalization of \deltafull\ to cases with multiple UV symmetries. 

Note that for $\CN=2$-preserving theories, we typically cannot use 't Hooft anomaly matching since we usually do not have an appropriate candidate superconformal $R$ symmetry. However, in conformal perturbation theory, we formally find a $U$ multiplet that satisfies
\eqn\Ntwocond{
\langle U(x)J_a(0)\rangle=0~,
}
for all preserved $\CN=2$ flavor symmetries, $J_a$ (these are symmetries that commute with the $\CN=2$ superconformal algebra). In $\CN=1$ theories, when we work in conformal perturbation theory, there is a natural $U$ operator that satisfies \Ntwocond\ and differs from the one determined by $a$-maximization by at most higher-order perturbative corrections.

In general, we should also include potential accidental symmetries---broken symmetries of the UV that become conserved in the IR---when we compute $a_{IR}$ and attempt to establish bounds on $\delta a$ (see \BuicanTY\ for a discussion of various subtleties and caveats; see also the interesting discussion in \ErkalSH).\foot{Our definition of accidental symmetry here does not include conserved symmetries of the UV theory that are broken by relevant deformations that decay to zero in the IR and do not change the course of the RG flow.} Typically, this is a difficult task. In the case of $\CN=2$ RG flows, we will show that under certain reasonable assumptions, accidental symmetries must descend from badly broken symmetries of the UV theory (note that the topological twisting technique used in \ShapereZF\ can also be very useful for many $\CN=2$ theories; see also the techniques in \XieJC\ and \AharonyDJ).

\newsec{$\CN=1$ theories}
Let us first discuss the case of $\CN=1$ RG flows. For such theories, it is straightforward to check that $d$ and $\delta a$ can be arbitrarily small. In other words, given any real number $r>0$, there exists an $\CN=1$ RG flow with $d, \delta a<r$.

By the discussion in the previous section, we should study RG flows that can be described at leading order in conformal perturbation theory. In $\CN=1$ SUSY there is a rich set of such RG flows. However, as we alluded to above, the simplest class of these theories---RG flows from asymptotically free fixed points to weakly interacting Banks-Zaks (BZ) theories in the IR---do not have arbitrarily small $d$ or $\delta a$. Formally, the reason that these quantities cannot be made arbitrarily small is that the Zamolodchikov metric is singular when we take the gauge coupling, $g$, to zero (see also the recent discussion in \KomargodskiXV) and compensates the small beta function---heuristically, the reason is that even though these flows are parametrically weakly coupled, the free UV fixed point is infinitely far away.

For example, consider $SU(N_c)$ SQCD with $N_f$ flavors and $N_c, N_f\to\infty$ with $x\equiv{N_c\over N_f}={1\over3}(1+\epsilon)$ fixed ($\epsilon\ll1$). In this regime, the theory is governed by the following beta function
\eqn\twoloop{
\beta_g=-{g^3\over16\pi^2}\left(3N x-N-{1\over8\pi^2}N^2xg^2\right)+\cdot\cdot\cdot~,
}
where $N=N_f$ and the interplay between the one and two-loop beta functions results in a parametrically small IR 't Hooft coupling, $\lambda_*\sim\epsilon$. At leading order, $\delta a$ is just
\eqn\lcha{
\delta a={16N^2x^2\over3}\int {dg\over g^2}\beta_g={2N^2\epsilon^2\over9}+\cdot\cdot\cdot~.
}
Since $\epsilon\sim N^{-1}$, we see that $\delta a\sim\CO(1)$ cannot be made arbitrarily small (note that $d\sim d_0=\infty$). The reason there is a lower bound on this quantity is that the factor of $g^{-2}$ arising from the Zamolodchikov metric enhances the UV contribution to $\delta a$.\foot{A more natural way to write \lcha\ is in terms of the holomorphic coupling, $\tau={4\pi i\over g^2}+{\Theta\over2\pi}$. Working in this variable, we find $\delta a\sim(N^2x^2-1)\int {d({\rm Im}\ \tau)\over ({\rm Im}\ \tau)^2}\beta_{{\rm Im}\tau}$, where the Zamolodchikov metric is now $g_{\tau\bar\tau}\sim({\rm Im}\ \tau)^{-2}$ \GreenDA.}

Clearly, to find an example of a class of RG flows with arbitrarily small $d, \delta a$, we should consider theories in which the Zamolodchikov metric is well-behaved. Furthermore, our discussion in the previous section motivates us to look for RG flows initiated by almost marginal deformations. To that end, consider the following weakly coupled class of SCFTs: the interacting fixed points of $SU(N_c)$ adjoint SQCD with $N_f$ flavors and vanishing superpotential in the limit $N_c, N_f\to\infty$ holding $x={1\over2}(1+\epsilon)$ fixed (for $\epsilon\ll1$). Let us further suppose that we add a free singlet chiral superfield, $\varphi$, to the UV theory. Next, let us deform the theory by turning on
\eqn\defasqcd{
\delta W=\lambda\varphi\Tr\ X^2~,
}
where $X$ is the adjoint superfield. Using $a$-maximization \IntriligatorJJ, we find that $D(\varphi\Tr\ X^2)=3-\epsilon+\cdot\cdot\cdot$, and so this deformation is slightly relevant and takes the theory to an IR fixed point with $\tilde R_{IR}(\varphi\Tr\ X^2)=3$ (note again that we have taken $N_f=N$). Computing $\delta a$, we find
\eqn\deltavsti{
\delta a={4\over3}\epsilon^2+\cdot\cdot\cdot\ll1~.
}
Clearly $\delta a$ (and $d\sim d_0$) can be made arbitrarily small as we take $\epsilon\sim N^{-1}\to0$ (and, since the flow is $R$-symmetric, $\tau_{U}\sim{3\over4}\delta a$ can be made arbitrarily small as well). Note that in the limit of large $N$, $\delta a$ is independent of $N$. The reason for this is simple: the leading quantum effects of the RG flow are captured by the change in the dimension of the $\varphi$ field (similar effects were noted in \GreenNQA). Since this operator is a singlet, it doesn't know about the gauge theory data (at leading order), and hence we expect $\delta a$ (and $d\sim d_0$) to be independent of $N$ when $N\gg1$.

We can also find examples of RG flows with arbitrarily small $\delta a$ that are fully interacting (i.e., no free fields anywhere along the RG flow) and strongly coupled. Indeed, consider the interacting fixed point of $SU(N_c)$ SQCD with $N_f$ flavors in the limit $N_c, N_f\to\infty$ with $x={1\over2}(1+\epsilon)$ (let us again set $N=N_f$). Since this theory has no adjoint superfield, it is very strongly coupled (in both the electric and magnetic descriptions). We can then turn on the following deformation
\eqn\Wsqcd{
\delta W=\lambda(Q_1\tilde Q_1)^2~.
}
Since $\tilde R(Q_i)=\tilde R(\tilde Q_i)={1\over2}(1-\epsilon)$, \Wsqcd\ is relevant, with dimension $3(1-\epsilon)$, and initiates an RG flow to an IR fixed point with $\tilde R_{IR}(Q_1)=\tilde R_{IR}(\tilde Q_1)={1\over2}$ with
\eqn\sqcdda{
\delta a={9N\over8}\epsilon^2+\cdot\cdot\cdot\ll1~.
}
Again we see that $\delta a$ can be made arbitrarily small as we take $\epsilon\sim N^{-1}\to0$ (and again, since the flow is $R$-symmetric, $\tau_{U}\sim{3\over4}\delta a$ can be made arbitrarily small as well). Note that at large $N$, $\delta a$ depends linearly on $N$. This dependence reflects the fact that we are (at leading order) changing the anomalous dimension of the $\CO(2N)$ degrees of freedom in $Q_1, \tilde Q^1$, while the remaining degrees of freedom are essentially inert.

\newsec{RG flows from $\CN=2$ SCFTs}
In this section we will discuss the case of RG flows emanating from an $\CN=2$ SCFT. We will consider both RG flows that preserve the full $\CN=2$ as well as those that break $\CN=2\to\CN=1$ explicitly.\foot{For a recent discussion on spontaneous breaking of $\CN=2$ SUSY, see \AntoniadisNJ.} We begin by recalling some basic facts about $\CN=2$ SCFTs.

\subsec{$\CN=2$ SCFTs}
An $\CN=2$ SCFT has an $SU(2)_R\times U(1)_R$ $R$ symmetry (if the SCFT consists of multiple decoupled sectors, then there are independent $SU(2)_R\times U(1)_R$ symmetries acting on the various subsectors). The corresponding currents reside in a real multiplet of dimension two and $SU(2)_R$ spin zero satisfying
\eqn\Ntwomult{
D^{(ij)}\CJ=0~,
}
where $i, j=1,2$ are $SU(2)_R$ indices, and the $D_{\alpha}^i$ are the chiral superspace derivatives. One particularly important operator in the $\CJ$ multiplet is the primary, $J=\CJ|$ (here the symbol \lq\lq $|$" indicates that we set the $\CN=2$ Grassman coordinates to zero). This operator is the moment map for the linear combination of $SU(2)_R\times U(1)_R$ that leaves the manifest $\CN=1$ superspace invariant (i.e., this linear combination is a flavor symmetry from the point of view of the manifest $\CN=1$ SUSY). The corresponding symmetry current, $J_{\mu}$, sits in the $\theta^1_{\alpha}\bar\theta_{1\dot\alpha}$ component of the $\CJ$ multiplet and satisfies
\eqn\Jmudef{
J_{\mu}=R_{\mu}^{\CN=2}-2I_{3\mu}~,
}
where $R_{\mu}^{\CN=2}$ is the $\CN=2$ superconformal $U(1)_R$ current, and $I_{3\mu}$ is the current corresponding to the $I_3\subset SU(2)_R$ symmetry. The superspace coordinates have charges $R_{\CN=2}(\theta_1)=R_{\CN=2}(\theta_2)=1$ and $I_3(\theta^1)=-I_3(\theta^2)={1\over2}$ under $U(1)_R$ and $I_3$ respectively (therefore, these coordinates have charges $J(\theta^1)=0$, $J(\theta^2)=2$ under $J$).

We can also consider $\CN=2$ flavor symmetry currents\foot{Below, we will also briefly discuss spin one currents that reside in multiplets with higher spin currents, but these multiplets will not play a significant role in our analysis.} (i.e., currents for symmetries that commute with the $\CN=2$ superconformal algebra). These operators reside in dimension two multiplets of $SU(2)_R$ spin one, $L^{ij}$, satisfying
\eqn\Lij{
L^{\dagger ij}=\epsilon_{ik}\epsilon_{jl}L^{kl}~, \ \ \ L^{ij}=L^{ji}~, \ \ \ D_{\alpha}^{(i}L^{jk)}=\bar D_{\dot\alpha}^{(i}L^{jk)}=0~.
}
From the last equation (and using the fact that $\bar D^2_{\dot\alpha}\sim \bar D_{1\dot\alpha}$ is the anti-chiral covariant derivative for the $\CN=1$ manifest superspace), we see that $L^{22}|=\mu$ is an $\CN=1$ chiral operator (this operator is often referred to as the \lq\lq holomorphic moment map" of the symmetry). On the other hand, we see that $iL^{12}|=J$ is the (real) moment map of the symmetry, since $\left[\left(Q^1\right)^2,J\right]=\left[\left(\bar Q_1\right)^2,J\right]=0$.\foot{We can make this discussion more concrete by considering the $U(1)$ flavor symmetry rotating a free hypermultiplet, $(Q, \tilde Q^{\dagger})$, by a common phase. In this case, $J\sim Q^{\dagger}Q-\tilde Q^{\dagger}\tilde Q$ and $\mu\sim Q\tilde Q$. Weakly gauging the symmetry we find that the K\"ahler potential includes a term $\delta K\sim VJ$ and the superpotential contains the $\CN=2$ companion term $\delta W\sim\Phi\mu$ where $(V, \Phi)$ is the $\CN=2$ vector multiplet (with $\Phi$ the component $\CN=1$ adjoint chiral superfield).}

Another important aspect of $\CN=2$ SCFTs that we will use below are the relations imposed by unitarity and the superconformal algebra on the dimensions, $R$ symmetry quantum numbers, and Lorentz spins of various short and semi-short representations (see, e.g., \refs{\DobrevQV\DolanZH\MinwallaKA-\GaddeUV} for additional details).
One particularly interesting class of operators for us below will be the operators satisfying $\bar Q_{\dot\alpha}^{(i}\CO^{i_1\cdot\cdot\cdot i_N)}_{\alpha_1\cdot\cdot\cdot\alpha_{2j}}=0$ with $Q_{\alpha}^{(i}\CO^{i_1\cdot\cdot\cdot i_N)}_{\alpha_1\cdot\cdot\cdot\alpha_{2j}}\ne0$ (in this case, $\CO^{11\cdot\cdot\cdot1}_{\alpha_1\cdot\cdot\cdot\alpha_{2j}}$ is  annihilated by $\bar Q_{2\dot\alpha}$). These operators satisfy
\eqn\chiralunit{
D=2j_R+{1\over2}R_{\CN=2}, \ \ \ R_{\CN=2}\ge2(j+1)~.
}
This class of operators includes the $SU(2)_R$ spin zero chiral operators that are annihilated by $\bar Q_{i\dot\alpha}$. The conjugate operators satisfy (in this case, $\CO^{\dagger 11\cdot\cdot\cdot1}_{\bar\alpha_1\cdot\cdot\cdot\bar\alpha_{2\bar j}}$, is annihilated by $Q^1_{\alpha}$)
\eqn\chiraluniti{
D=2j_R-{1\over2}R_{\CN=2}, \ \ \ R_{\CN=2}\le-2(\bar j+1)~.
}
Other interesting operators include those defined by the conditions $\bar Q_{\dot\alpha}^{(i}\CO^{i_1\cdot\cdot\cdot i_N)}=Q_{\alpha}^{(i}\CO^{i_1\cdot\cdot\cdot i_N)}=0$. This class of operators includes the $SU(2)_R$ spin one-half hypermultiplet and the $SU(2)_R$ spin one conserved current multiplet (see \Lij; the operator $\CO^{11\cdot\cdot\cdot1}$ is annihilated by both $Q^1_{\alpha}$ and $\bar Q_{2\dot\alpha}$). These operators satisfy
\eqn\chiralunitii{
D=2j_R, \ \ \ j_{R}\ge{1\over2}, \ \ \ R_{\CN=2}=0~.
}
Finally, the multiplet that includes the $SU(2)_R\times U(1)_R$ currents and stress tensor satisfies
\eqn\stressunit{
D=2, \ \ \ j_R=R_{\CN=2}=0~,
}
while higher spin currents sit in multiplets with $D=2+j+\bar j>2$.

\subsec{$\CN=2$ RG flows}
Before proceeding to discuss relevant deformations of $\CN=2$ SCFTs, we note in passing that there are no direct $\CN=2$ analogs of the $\CN=1$ BZ flows. Indeed, consider $\CN=2$ $SU(N_c)$ SQCD with $N_f$ flavors and ${N_c\over N_f}={1\over2}(1+\epsilon)$ fixed in the limit $N_c, N_f\to\infty$ with $\epsilon\ll1$. If we turn on some non-zero gauge coupling, we find that the perturbative beta function is saturated at one loop (see \SeibergUR\ and references therein) and so, unlike in the $\CN=1$ case around \twoloop, there is no IR fixed point at small but finite gauge coupling.\foot{Indeed, the anomalous dimensions for the hypermultiplets vanish since the corresponding mass terms are related to conserved flavor currents by $\CN=2$ SUSY as in \Lij. We can also check that there are no fixed points at small coupling since any putative $R$ symmetry would assign charge zero to the adjoint, $\Phi$. Therefore, gauge invariant monomials like $\Tr\ \Phi^N$ would also have zero superconformal $R$ charge---thereby violating unitarity. Furthermore, under this putative $R$ symmetry, we would have $a_{IR}=c_{IR}=0$, which would be inconsistent with an interacting IR fixed point.}

Since we wish to understand if it is possible to construct $\CN=2$ RG flows with $d$ and $\delta a$ parametrically small, it is natural to study $\CN=2$ SCFTs with almost marginal relevant deformations.\foot{Recall from the discussion around \twoloop\ that RG flows with $d$ and $\delta a$ parametrically small should have a well-defined Zamolodchikov metric. Therefore, we do not consider the possibility of adding asymptotically free gauge fields to the UV theory.} Note, however, that we will consider $\CN=2$ theories with more general relevant deformations at the end of this section. We begin by discussing the RG flow in the vicinity of the UV fixed point and studying $\delta a$.

Choosing some manifest $\CN=1\subset\CN=2$ SUSY, unitarity requires that our relevant deformation resides in the superpotential. Furthermore, the $\CN=2$ superconformal algebra combined with unitarity restricts our relevant deformations to be of the types with dimensions given in \chiralunit\ (with $j_R=0, 1/2$) or their $Q_{\alpha}^2$ and $(Q_{\alpha}^2)^2$ descendants (as long as $j_R=0$).\foot{Note that the operators described around \chiralunitii, while relevant for $j_R\le1$, have integrally quantized dimensions and are therefore not almost marginal; however, we will discuss these deformations at the end of this section when we relax the condition on the dimension of the relevant deformation.} More precisely, we turn on
\eqn\WdefNone{
\delta W=\lambda^i\CO_i~, \ \ \ D(\CO_i)=3-\epsilon_i~,
}
where each $\CO_i$ must be characterized by one of the following sets of properties:

\smallskip
\noindent
{\bf (i)} $\CO_i=\left\{Q^{2\alpha},\left[Q^2_{\alpha},\hat\CO_i\right]\right\}|_{\theta^2,\bar\theta_2}$, with $\hat\CO_i$ an $\CN=2$ primary of $SU(2)_R$ spin zero and $R(\hat\CO_{i})=2(2-\epsilon_i)$ as in \chiralunit; therefore, $J(\CO_i)=-2\epsilon_i$ and $R(\CO_i)=2-2\epsilon_i$, or

\smallskip
\noindent
{\bf (ii)} $\CO_i=\left[Q^{2\alpha},\hat\CO_{i\alpha}\right]|_{\theta^2,\bar\theta_2}$, with $\hat\CO_{i\alpha}$ an $\CN=2$ primary of $SU(2)_R$ spin zero and $R(\hat\CO_{i\alpha})=5-2\epsilon_i$ as in \chiralunit; therefore, $J(\CO_i)=3-2\epsilon_i$ and $R(\CO_i)=4-2\epsilon_i$, or

\smallskip
\noindent
{\bf (iii)} $\CO_i$ is an $\CN=2$ primary with $SU(2)_R$ spin zero, $R(\CO_i)=2(3-\epsilon_i)$, and $J(\CO_i)=2(3-\epsilon_i)$ as in \chiralunit, or 

\smallskip
\noindent
{\bf (iv)} $\CO_i$ is an $\CN=2$ primary with $SU(2)_R$ spin one-half (more precisely, the component annihilated by $\bar Q_{1\dot\alpha}$), $R(\CO_i)=2(2-\epsilon_i)$, and $J(\CO_i)=3-2\epsilon_i$ as in \chiralunit.
\smallskip
\noindent
The deformations of class {\bf(i)} are the $\CN=2$-preserving prepotential deformations. The deformations of type {\bf (ii)-(iv)} preserve an $\CN=1\subset\CN=2$ sub algebra. For example, the deformations of type {\bf (iii)} are similar in nature to the $\CN=2\to\CN=1$ breaking adjoint mass deformation, $\delta W=m\Tr\Phi^2$ (see \TachikawaTT\ for an interesting analysis of this term). The only difference here is that we are considering almost marginal deformations, although we will relax this condition below.

Let us now suppose that we deform our theory by turning on operators exclusively of type {\bf (i)}, thus perserving $\CN=2$.\foot{We assume that the relevant deformation does not lead to spontaneous breaking of $\CN=2$ SUSY. In fact, since this deformation preserves $SU(2)_R$, we can invoke the results of \AntoniadisNJ\ to conclude that SUSY breaking is unlikely.} In more covariant notation, this deformation is just
\eqn\prepot{
\delta\CL\sim\int d^2\theta^1d^2\theta^2\lambda^i\hat\CO_i+\rm{h.c.}~,
}
where the integration is over the chiral half of $\CN=2$ superspace and $D(\hat\CO_i)={1\over2}R(\hat\CO_i)=2-\epsilon_i$ (i.e., we use \chiralunit\ with $j=j_R=0$). This deformation clearly breaks the $J$ symmetry of the UV SCFT. As discussed above, we could consider a UV theory with several decoupled sectors, in which case we would have multiple independent $J$ currents. However, the deformation in \prepot\ cannot couple these sectors, because this would require a composite $\hat\CO_i$ operator built out of gauge invariant operators from the various subsectors (unitarity requires each such operator to have dimension at least one, which means that the dimension of $\hat\CO_i$ would be at least two). Therefore, without loss of generality, we assume that the UV theory has a single $J$ current.

Now, since the operators in \prepot\ are singlets under $SU(2)_R$, we see that they are uncharged under any flavor symmetries, $F_{UV}$, of the UV SCFT. Indeed, we find that
\eqn\chiralflavor{
\langle\hat\CO_i(x)\hat\CO^{\dagger}_{\bar j}(y)J_{F_{UV}}(0)\rangle=0~, \ \ \ \hat\CO_i(x)\hat\CO_{\bar j}^{\dagger}(0)\not\supset {T^a_{i\bar j}\over x^{2-2\epsilon}}J_{F_{UV}a}~,
}
where $J_{F_{UV}a}$ is the real moment map for the generator $t_a$ of the $F_{UV}$ symmetry (see the discussion around \OPEs\ and \Tibarjdef). The expressions in \chiralflavor\ follow from the fact that $J_{F_{UV}a}$ transforms as the $\sigma_3$ part of an $SU(2)_R$ triplet of fields (these statements are just the straightforward generalizations of the fact that the $\Tr\Phi^k$ Coulomb branch operators are uncharged under flavor symmetries).

As a result, we see that the deformation \prepot\ preserves the $SU(2)_R$ symmetry as well as the full $F_{UV}$ flavor group.\foot{Since the $F_{UV}$ symmetries are non-chiral, i.e., $\Tr J_{F_{UV}}=\Tr J_{F_{UV}}^3=0$, it is in principle possible that $F_{UV}\to0$ in the IR and decouples from the low energy physics. Note that $\Tr J_{F_{UV}}=0$ follows from the fact that $J$ cannot appear in the $J(x)J_F(0)$ OPE, because it has the wrong $SU(2)_R$ quantum numbers. Similarly, $\Tr J_{F_{UV}}^3=0$ follows from noting that $J^I_{F_{UV}}(x)J_{F_{UV}}^J(0)=\cdot\cdot\cdot+\epsilon^{IJ}_{\ \ K}\cdot{\kappa\over x^2}J^K_{F_{UV}}+\cdot\cdot\cdot$. Since the real moment map, $J_F$, has $I=3$, it follows that $J_F$ does not appear in the $J_FJ_F$ OPE, and so the flavor symmetries cannot have cubic 't Hooft anomalies.} In particular, with respect to the $\CN=1$ superspace, $S_1$, parameterized by $\theta^1_{\alpha},\bar\theta_{1\dot\alpha}$, the deformation in \prepot\ preserves a $U(1)_R$ symmetry given by
\eqn\Rsymmpres{
R=\tilde R_{UV}^{\CN=1}-{1\over3}J=2I_3~, \ \ \ U=-{1\over2}J~,
}
where $\tilde R_{UV}^{\CN=1}={1\over3}R_{UV}^{\CN=2}+{4\over3}I_3$ is the UV superconformal $R$ symmetry with respect to $S_1$. However, it is clear that \Rsymmpres\ has the wrong $SU(2)_R$ quantum numbers to flow to the IR superconformal $R$ symmetry (with respect to $S_1$).\foot{In fact, we reach the same conclusion even if we consider a more general linear combination of \Rsymmpres\ with any of the flavor symmetries of the UV theory (since the corresponding currents are $SU(2)_R$ singlets). This statement is not surprising since these symmetries are non-chiral and hence do not mix with the superconformal $R$ current \IntriligatorJJ.}
Actually, we can be more careful and distinguish between two possibilities:

\smallskip
\noindent
{\bf 1.} The UV theory flows to a gapped phase (we do not know of any such examples).\foot{Of course, the massive free hypermultiplet flows to a gapped phase in the IR. However, this theory has no $\hat O_i$ operators.}

\smallskip
\noindent
{\bf 2.} The IR theory is not gapped (we would generally expect that the IR theory has a moduli space).

\smallskip
\noindent
In the hypothetical first case, \Rsymmpres\ correctly implies that $a_{IR}=c_{IR}=0$. It then follows that $\delta a=a_{UV}$  and $\delta c=c_{UV}$ cannot be parametrically small. There are at least two ways to see this. One way is more specific to (certain) $\CN=2$ theories, while the other is more general. The more $\CN=2$-specific method is the following. For UV $\CN=2$ theories in which the formula \refs{\ArgyresTQ, \ShapereZF}
\eqn\Rcoulombi{
2a-c={1\over4}\sum_i(R_{\CN=2}(\hat\CO_i)-1)~, \ \ \ R(\hat\CO_i)>2 \ \ \ \forall i~,
}
holds (note that in \Rcoulombi\ we sum over all $SU(2)_R$ singlet scalar chiral primaries of dimension larger than one; in \ShapereZF, this equality was demonstrated for any $\CN=2$ SCFT that can be embedded in an $\CN=2$ gauge theory), we see that the existence of the $\hat\CO_i$ in \prepot\ implies 
\eqn\deltaa{
2a_{UV}-c_{UV}\ge\sum_i{1\over4}(3-2\epsilon_i)>{N_{\CO}\over4}\ge{1\over4}~.
}
In arriving at the first inequality we have substituted the $R_{\CN=2}$ charges of the $\hat\CO_i$ into \Rcoulombi, while in the second inequality we have used the fact that each of the $N_{\CO}\ge1$ $\hat\CO_i$ operators has $R_{\CN=2}(\CO_i)>2$. Next, recall that unitarity implies $c_{UV}>0$ . Therefore, in any hypothetical theory with a relevant prepotential deformation that takes the theory to a gapped phase (and for which \Rcoulombi\ applies), we have
\eqn\aboundNtwo{
\delta a>{1\over8}~.
}

A more general argument showing that $\delta a$ is bounded positively from below can be formulated as follows: the conformal bootstrap results of \refs{\PolandWG, \RattazziGJ, \PolandEY} show that in any CFT (certainly ones with scalar primaries of dimension $2-\epsilon_i$), the $c$ central charge cannot be arbitrarily small (note that unitarity alone only shows that $c>0$ and therefore does not rule out $c$ parametrically smaller than one). In particular, these authors showed that in a general CFT with a dimension $d$ scalar primary
\eqn\CFTbounds{
c\ge f(d)>0~,
}
where $f$ is a universal function of the dimension, $d$ (which is finite around $d=2$). Our $\hat\CO_i$ have $d=2-\epsilon_i$. If we assume that these are the lowest dimensional scalar operators in the theory, it turns out that \RattazziGJ\foot{Our normalization of $c$ differs from the one in \RattazziGJ; we have $c^{\rm here}={1\over15}c^{\rm there}$.}
\eqn\bootstrapf{
c_{UV}\gtrsim{1\over60}~.
}
Now, recalling the central charge bounds of Hofman and Maldacena for $\CN=2$ theories \HofmanAR
\eqn\HMNtwo{
{1\over2}\le{a\over c}\le{5\over4}~,
}
we see that
\eqn\aUVbound{
\delta a\gtrsim{1\over120}~.
}
This bound is weaker than the one derived by using the formula in \Rcoulombi\ along with unitarity, but the reasoning is more general. Note that if our $\hat\CO_i$ are not the lowest dimension scalar operators in the theory, then we expect the bound in \aUVbound\ to become stronger (using the reasoning in \RattazziGJ\ applied to these lower-dimension operators).

Let us now consider the second (much more likely) case, namely that the IR theory is not gapped (or, at least, the sector of the theory with the $\hat\CO_i$ is not gapped). If the IR theory is not described in terms of the same degrees of freedom as the UV theory (this condition includes simple cases where we give mass to free fields), then (as we discussed in section two) it must be the case that $d$ is bounded positively from below (and similarly for $\delta a$ under the assumption of a non-fine-tuned decreasing $a$-function; these results also follow if the UV or the IR has an ill-defined Zamolodchikov metric). On the other hand, if these conditions are not satisfied, then the $a$-theorem implies that some accidental symmetries must emerge and mix with $R$ in \Rsymmpres\ to produce the correct IR superconformal $R$ current. The currents corresponding to these accidental symmetries descend from UV vector operators, $\CO_{\mu}$, of dimension $D=3+\gamma>3$. In the appendix, we argue that approximately conserved currents of the UV SCFT---i.e. $\CO_{\mu}$ operators with $\gamma\ll1$---cannot mix with the IR superconformal $R$ current under these conditions. Therefore, the accidental symmetries must emerge from badly broken UV symmetries with $\gamma$ bounded finitely from below. As a result, $d$ cannot be arbitrarily small in this case either. Finally, under the assumption that a non fine-tuned $a$ function exists, $\delta a$ should be bounded positively from below as well.

Another way to see that the description of the system must be more complicated is to examine the RG flow at leading order in conformal perturbation theory. Since the Zamolodchikov metrics, $g_{i\bar j}$, of these theories are non-singular, an IR fixed point in the neighborhood of our UV theory would yield
\eqn\deltaaleading{
\delta a\sim{4\over3}\tau_{U}={1\over3}\tau_J~,
}
parametrically small in the $\epsilon_i$, where we have used \confperta, \Rsymmpres, and \Ntwocond.\foot{Indeed, note that 
\eqn\ortho{
\langle U(x)J_{F_{UV}}(0)\rangle=-{1\over2}\langle J(x)J_{F_{UV}}(0)\rangle=0~,
}
where $J_{F_{UV}}$ is the real moment map of some element in the UV flavor symmetry group. The last equation follows from the fact that $J$ is an $SU(2)_R$ singlet, while $J_{F_{UV}}$ is part of an $SU(2)_R$ triplet.} Now, since $J$ is in the same multiplet as the stress tensor, $\tau_J\sim c_{UV}$. More precisely,
\eqn\taujc{\eqalign{
\tau_J&=-3\Tr\ \left({1\over3}R_{\CN=2}^{UV}+{4\over3}I_3^{UV}\right)\left(R_{\CN=2}^{UV}-2I_3^{UV}\right)^2=-\left(\Tr\ \left(R_{\CN=2}^{UV}\right)^3-12\Tr\ R_{\CN=2}^{UV}\left(I_3^{UV}\right)^2\right)\cr&=-({9\over2}(a_{UV}-c_{UV})-{9\over8}(4a_{UV}-2c_{UV}))={9\over4}\cdot c_{UV}~.
}}
Therefore, we see that at leading order in conformal perturbation theory
\eqn\deltaavsc{
\delta a={3\over4}\cdot c_{UV}~.
}
From \deltaavsc, we see that if $\delta a$ is parametrically small, then so too is $c_{UV}$---this is a contradiction. Indeed, we can again use the formula for $2a-c$ in \Rcoulombi\ to show that $2a_{UV}-c_{UV}>{1\over4}$. Clearly if $c_{UV}\ll1$, then the bounds of Hofman and Maldacena \HofmanAR\ are badly violated. Alternatively, we can again use the bootstrap results of \refs{\RattazziGJ, \PolandWG, \PolandEY} to show that $c_{UV}$ cannot be parametrically small.

Given these various obstructions, it seems unlikely that the $\CN=2$ RG flow allows theories in which $\delta a$ or $d$ can be parametrically small. Although we have not discussed deformations by more strongly relevant operators, based on our above discussion, we do not expect that we can produce small $\delta a$ in this case either. We will return to such deformations in the next subsection.

Let us now discuss the other allowed relevant deformations of our $\CN=2$ UV SCFT. We will also begin by assuming that we have unique set of $SU(2)_R\times U(1)_R$ currents (and a unique stress tensor). We will later relax this requirement. Note that in cases {\bf(ii)} or {\bf(iii)}, our deforming operators must be singlets under the $\CN=2$ flavor symmetry group, $F_{UV}$, by logic similar to the reasoning used around \chiralflavor. Only if the deforming operators are $\CN=2$ primaries of $SU(2)_R$ spin one-half (case {\bf (iv)} above) can the deformation be charged under $F_{UV}$. In Lagrangian theories, deformations of type {\bf(iv)} exist only when there are free gauge-invariant hypermultiplets (or if the theory has $\CN=4$ SUSY; we will discuss $\CN=4$ theories in the next section). More generally, under the assumption of having just a single set of $SU(2)_R\times U(1)_R$ currents and a single stress tensor (with at most $\CN=2$ SUSY), we are not aware of any examples of such operators. Therefore, we will focus on the remaining cases {\bf (ii)} and {\bf (iii)} in which the $\CO_i$ in \WdefNone\ are singlets under $F_{UV}$.

From the charges of the $\CO_i$ under $R$ and $J$, we see that for generic deformations of type {\bf (ii)} and / or {\bf (iii)}, \WdefNone\ doesn't preserve an $R$ symmetry. Since the IR endpoint of the flow must have a superconformal $R$ symmetry, $\tilde R_{IR}$, it follows from the proof in the appendix that some of the couplings in \WdefNone\ must flow to zero if we are to remain in a sufficiently small neighborhood of $\CT_{UV}$. As a result, nearby fixed points can always be reached by considering RG flows generated by adding a subset of the $\CO_i$ that preserve a common $R$-symmetry to the superpotential.

For simplicity, let us consider an RG flow with $\delta W=\lambda^i\CO_i$ of type {\bf (iii)} above (other cases can be treated similarly). In this case, we have an $R$ symmetry with $U(\CO_i)=\epsilon$ and $U={\epsilon\over2(3-\epsilon)}J$, i.e., the $R$ symmetry current is just $\hat R_{\mu}=\tilde R_{\mu}^{UV}+{\epsilon\over3(3-\epsilon)}J$, where $\tilde R_{\mu}^{UV}$ is the superconformal $R$ symmetry current with respect to a chosen $\CN=1\subset\CN=2$ sub-algebra.

From the discussion in section two, we see that, at leading order,
\eqn\deltaaleading{
\delta a\sim{4\over3}\tau_{U}={\epsilon^2\over3(3-\epsilon)^2}\tau_J={3\epsilon^2\over4(3-\epsilon)^2}c_{UV}~.
}
In the second and third equalities we have again used \ortho\ and \taujc\ respectively. Note that the factor of $\epsilon^2$ saves us from the various obstructions to finding $\delta a$ parametrically small that we found in the $\CN=2$-preserving RG flows. Instead, we find various consistency conditions on the UV central charges in theories with $\delta a\ll1$.

In particular, we see that in order for there to be an $\CN=2\to\CN=1$ breaking RG flow that is parametrically short, it must be the case that
\eqn\UVcentral{
a_{UV}, \ c_{UV}\sim\CO(\epsilon^{-N}), \ \ \ N<2~.
}
As a result, we see that if the UV central charges are too large (relative to the inverse of the deviation from marginality of the relevant deformation) then $\delta a$ will not be small and an $\CN=1$ SCFT will not be \lq\lq parametrically close" to the UV $\CN=2$ SCFT ($d_0$ will be parametrically small if $1<N$).\foot{This statement is still true if we take into account the full perturbative corrections in the neighborhood of the UV SCFT. In this case, \deltaaleading\ becomes
\eqn\deltaamodified{
\delta a={\epsilon^2(9c_{UV}-4\epsilon\cdot a_{UV})\over4(3-\epsilon)^3}~,
}
which is still small if and only if \UVcentral\ is satisfied (note that the Hofman-Maldacena bounds prevent the $c_{UV}$ and $a_{UV}$ terms from approximately canceling in \deltaamodified).} In particular, there can be no analog of the flow considered in \Wsqcd, because in that flow we have $a_{UV}, c_{UV}\sim\CO(\epsilon^{-2})$. Note that the flow in \Wsqcd\ has parametrically small $\delta a$ because we only excite $\CO(\epsilon^{-1})$ degrees of freedom at leading order. On the other hand, by comparing \deltaaleading\ to \deltavsti\ and \sqcdda, we see that in flowing from an $\CN=2$ SCFT to a parametrically close $\CN=1$ SCFT we excite of order all the degrees of freedom in the $\CN=2$ SCFT (since we excite of order the number of degrees of freedom that contribute to $c_{UV}$).

If, on the other hand, the UV theory has two or more sets of $SU(2)_R\times U(1)_R$ currents and stress tensors, then we do not generally find simple relations of the form \deltaavsc. For example, suppose we have some UV SCFT, $\CT$, and some free decoupled hypermultiplet, $(Q, \tilde Q^{\dagger})$. If the theory has a chiral primary of dimension $2-\epsilon$, $\hat\CO$, then we can for example couple it to $Q$ via
\eqn\deltaW{
\delta W=\lambda\hat\CO Q~.
}
The deformation in \deltaW\ is of type {\bf (iv)} since it has $SU(2)_R$ charge one-half and $U(1)_R$ charge $2(2-\epsilon)$. Furthermore, it is clearly charged under the symmetry that rotates $Q$ by a phase. We find that at leading order in conformal perturbation theory
\eqn\deltaaQdec{
\delta a\sim{12\epsilon^2\cdot c_{UV}\over 9\cdot c_{UV}+16(2-\epsilon)^2}~,
}
where $c_{UV}$ is the central charge for the interacting part of the UV theory. Therefore, even if $c_{UV}\sim\CO(\epsilon^{-N})$ with $N\ge2$, we see that $\delta a\sim\epsilon^2$ can be parametrically small, and the flow can be described by leading order conformal perturbation theory. The reason that $\delta a$ can be parametrically small in this case is that, as in the $\CN=1$ theory of \defasqcd, the leading order quantum effects in this limit are captured by the change in the dimension of the free field.\foot{One can check that these statements still hold if we take into account higher-order quantum corrections.}

Note that if we can find examples of operators of type {\bf(iv)} in theories with a single stress tensor, then we expect that it is possible to find results similar to \deltaaQdec\ in such theories, and the bounds in \UVcentral\ will not necessarily hold. Finally, if we include deformations that are (marginally) irrelevant at $\CT_{UV}$ but that become relevant after we turn on our genuinely relevant deformations, then we again expect that we can find results similar to \deltaaQdec\ in theories with a single stress tensor (after all, we can add operators of the type \chiralunitii\ with $j_R={3\over2}$).

\subsec{More general relevant deformations: away from marginality}
In our discussion thus far we have assumed that the relevant deformations we turn on are almost marginal (since the corresponding RG flows are, via the discussion in section two, the only way to potentially find RG flows with $d\sim d_0$ arbitrarily small and are also natural candidates for theories with $\delta a$ arbitrarily small). Modulo some small assumptions, we can show that turning on more strongly relevant deformations does not dramatically alter our conclusions (let us emphasize that, in any case, $d$ will not be parametrically small in the RG flows of this section).

Indeed, the discussion around \Rsymmpres\ based on the $SU(2)_R$ quantum numbers of the preserved $R$ symmetry still holds for cases in which the prepotential deformations are highly relevant (since the identification $U=-{1\over2}J$ is independent of the dimension of the $\hat\CO_i$).  In fact, the discussion around \deltaavsc\ holds in these cases as well, and so does the reasoning in the appendix.\foot{We should note that turning on a relevant deformation may also force us to turn on various non-trivial vevs that break $J$. In our class of RG flows, we assume that turning on any such vevs is a sub-leading breaking of conformal invariance in the UV.}

One small subtlety in our discussion of $\CN=2$ RG flows is that if we consider highly relevant prepotential deformations, then we should also consider an $\CN=2$-preserving deformation of the form
\eqn\deltaW{
\delta W=m^a\mu_a~,
}
where the $\mu_a$ are some of the dimension two holomorphic moment maps for the UV flavor symmetry group (this is the generalization of adding a mass term for a hypermultiplet in a weakly coupled gauge theory).\foot{Note that we could also imagine turning on an (electric and / or magnetic) $\CN=2$ FI term. For example, we could take $\delta W=f\Phi$, where $\Phi$ is the $\CN=2$ partner of the $U(1)$ vector multiplet. This deformation explicitly breaks $SU(2)_R$ and could lead to spontaneous SUSY breaking. If $\CN=2$ SUSY is not broken, then the vacuum of the theory must shift, and the $U(1)$ symmetry will be Higgsed. However such a scenario requires turning on an additional (marginally) irrelevant deformation (the $U(1)$ gauge coupling), and we do not consider such theories in this paper.} For simplicity, we will assume that none of the $\mu_a$ become irrelevant in the IR.\foot{If some of the $\mu_a$ do become irrelevant in the IR, there are two cases to check. First, suppose the IR irrelevant $\mu_a$ play no role in the flow from $\CT_{UV}\to\CT_{IR}$. In this case, we can omit them from the UV deformation in \deltaW, and our discussion is not affected. On the other hand, if the IR irrelevant $\mu_a$ play an important role at short distances, we should not ignore them in our analysis, and using 't Hooft anomaly matching to compute $\delta a$ will typically be unreliable.}

Note that with deformations of the form \deltaW, we can couple UV SCFTs with multiple decoupled sectors. However, unitarity restricts this coupling to be a simple mass term for free hypermultiplets (i.e., $\delta W\sim mQ_1\tilde Q_2$). In this case, $\delta a\ge{4\over9}$. It remains to consider the (more interesting) case of couplings of the form \deltaW\ in a UV theory with a single sector.

One interesting phenomenon that can occur in this case is that if the deformations in \deltaW\ include non-abelian moment maps, then we can have broken $\CN=2$ flavor symmetries that contribute to $U$ (the $\mu$ operators for abelian symmetries are only charged under $J$). In this case, $\delta W$ can have (quantized) charges, $q_{F_i}$, under some non-abelian flavor symmetry generators, $T_{F_i}$ (with associated currents, $J_{F_i}$). We can then choose some particular broken symmetry current, $J_{F}$, that is orthogonal to all the preserved flavor symmetries of the theory. Denoting $J_F(\delta W)=q_F$, we can attempt to use $a$-maximization to compute the mixing of the conserved current $J'=J+{2\over q_F}J_F$ with $U$. Proceeding in this way, we find that our trial $R$ current is just $R_b=\tilde R_{UV}+{2\over3}\left(-{1\over2}J+bJ'\right)$, and that $a$-maximization selects
\eqn\bamax{
b={6q_F^2(2a_{UV}-c_{UV})-\sqrt{q_F^2\left(9q_F^2c_{UV}^2+48\tau_F(4a_{UV}-c_{UV})\right)}\over6(4a_{UV}-3c_{UV})q_F^2-32\tau_F}~,
}
where $\tau_F$ is the two-point function for $J_F$. Note that the Hofman-Maldacena bounds (along with unitarity in the guise of $\tau_F>0$) guarantee that $b$ is real (one can also check that $|b|<\infty$). 

However, in order to have an IR superconformal $R$ current with appropriate $\CN=2$ quantum numbers, we need $b={1\over2}$, and we can see from \bamax\ that this condition cannot be satisfied for $q_F<\infty$. Therefore, we are again in the situation of the previous subsection. Another possibility is that our relevant operator is related to the nilpotent deformations discussed in \HeckmanQV. There the authors concluded that such deformations break $\CN=2\to\CN=1$.\foot{In some cases, one can also have the opposite situation: an $\CN=1$ SCFT in the UV and an $\CN=2$ SCFT in the IR. In this scenario, \HeckmanQV\ uses the holographically derived results of \AharonyDJ.} 

In this case, we use \bamax\ to compute $\delta a=a_{UV}-a_{IR}$. This function doesn't have extrema at finite values of $q_F$. Instead, it slopes to zero as $q_F\to\pm\infty$ with
\eqn\deltaadiff{
\delta a\ge\delta a_{\rm min}={4\tau_F\over3q_F^2}~.
}
Although the final quantity in \deltaadiff\ has not, to our knowledge, been directly bounded from below (or even studied) in the literature, many similar lower bounds on current two-point functions to operator charge squared ratios exist in the conformal bootstrap literature (see, e.g., \refs{\PolandEY,\PolandWG} and the discussion around \bounds). It is therefore reasonable to assume that the quantity in \deltaadiff\ can be bounded universally from below using similar techniques (in particular, one would  need to generalize the bounds of \PolandEY\ on fundamental matter to adjoint matter). Note also that we have not been careful to check whether unitarity bounds are violated. Such violations indicate the presence of accidental symmetries, and we may run into problems when using 't Hooft anomaly matching to compute $\delta a$.

Let us now discuss other $\CN=2\to\CN=1$ breaking RG flows in the presence of more highly relevant deformations, $\delta W=\lambda^i\CO_i$. We can again attempt to use a visible $R$ symmetry and see if we can find $\delta a$ small in this case, since we can often find candidate superconformal $R$ currents that pass certain tests (keep in mind that $d_0$ will still not be small in this case). In order to find some candidate $R$ currents, we should not turn on $\CO_i$ of type {\bf (iii)} (or type {\bf(ii)}) and deformations of the type in \deltaW\ simultaneously since they do not preserve a common $R$ symmetry.\foot{Since the IR theory must have an $R$-symmetry, it may happen that some of the relevant UV operators become irrelevant and decay to zero in the IR. If these operators do not affect the flow, we may simply drop them from the UV superpotential deformation. On the other hand, if these IR irrelevant operators play a role at short distances, we will generically have accidental symmetries in the IR when these operators decay to zero, and we will typically not be able to use 't Hooft anomaly matching to compute $\delta a$.} Let us focus on deformations of type {\bf(iii)}. From \deltaamodified, we see that the full $\delta a$ cannot be small unless the dimensions of the deforming operators are either close to three (as in the discussion of the previous subsection) or if the dimensions are in the range $1<D\lesssim{6\over5}$.\foot{In the latter case, we might still think of the RG flow as exciting all the degrees of freedom in the theory since $U\sim J$.} Note that if $1<D\lesssim{6\over5}$, and the UV theory has any $\CN=2$ flavor symmetry, then we have been too quick in trusting \deltaamodified\ (in fact, this formula breaks down already for slightly larger $D$). Indeed, the holomorphic moment map operators, $\mu_a$, for the $\CN=2$ flavor symmetry violate the unitarity bound in the IR unless accidental symmetries emerge. In particular, under the $R$ symmetry discussed above \deltaaleading, we find that $\tilde R(\mu_a)\lesssim{1\over3}$. Of course, we are assuming that the $\mu_a$ do not flow to zero in the IR. Presumably, if this were the case, then $\delta a$ would be bounded from below (at least if the $\mu_a$ are holomorphic moment maps for non-abelian symmetries).\foot{Recall from the discussion in the introduction that if we can generalize the bounds of \PolandEY\ to arbitrary non-abelian symmetry groups and matter representations, then we can bound $\tau_{a}^{UV}>0$ positively from below and so we would expect that if $\mu_a$ flows to zero, then $\delta a$ is bounded positively from below as well (as long as we can write a non-fine-tuned $a$-function).}

As a result, by our above discussion and the reasoning in appendix A, we expect that $\delta a$ cannot be made parametrically small in such RG flows. In particular, we see that the central charge bounds in \UVcentral\ should be respected unless we start from a UV SCFT without $\CN=2$ flavor symmetry, and we deform it with operators, $\CO_i$, of $1<D\lesssim{6\over5}$. In this case, if we can tune $a_{UV}\sim{9\over4(3-D)}c_{UV}$, then it may be possible to find $\delta a$ parametrically small and to simultaneously violate the bounds in \UVcentral, although such a solution may still fail. For example, we should analyze other operators for unitarity bound violations or look for more subtle problems with the descriptions of these RG flows.

\subsec{Examples}
One natural way to obtain deformations of type {\bf(i)} or of type {\bf (iii)} is to consider Argyres-Douglas (AD) theories \refs{\ArgyresJJ\ArgyresXN\EguchiVU-\GaiottoJF} (see also the recent review in \GiacomelliTIA) and their generalizations \refs{\XieHS, \XieJC} (e.g., theories found by reducing the six-dimensional $A_{N-1}$ $(2,0)$ theory on a sphere with one irregular singularity and either zero or one additional regular singularities). Such theories are interesting because they have chiral primaries with rational dimensions.\foot{Theories found by reducing the $A_{N-1}$ $(2,0)$ theory on an arbitrary genus surface with only regular singularities \GaiottoWE\ yield integer scaling dimensions and are therefore only potentially useful if we wish to study RG flows with strongly relevant deformations.} Indeed, since the rationals are dense in the real numbers, we can also consider a certain limiting class of these theories and often find $\CN=2$ SCFTs with prepotential deformations of dimension $2-\epsilon$ and $\epsilon\ll1$ parametrically small (we can also often find $\CN=2\to\CN=1$ breaking superpotential deformations of dimension $3-\epsilon$).

For example, we can consider $\CN=2$ preserving RG flows of type {\bf (i)} between the $I_{2,2n}$ and $I_{2,2(n-1)}$ SCFTs of \XieJC\ (here $n$ is an integer; these SCFTs were originally studied in \EguchiVU\ and are the AD points of the $SU(2n)$ and the $SU(2n-2)$ gauge theory respectively). In this case, one finds an almost marginal operator, $\hat\CO_{I_{2,2n}}$, of dimension $D(\hat \CO_{I_{2,2n}})={2n\over n+1}\sim 2\left(1-{1\over n}\right)$ when $n\gg1$. For these RG flows $\delta a={16\over3}\left(1-{1\over n}+{1\over(1+n)}\right)\ge{8\over3}$, and $\delta a\to{16\over3}$ in the large $n$ limit.\foot{Our normalization for $a$ differs from the one in \XieJC\ by a factor of ${32\over 3}$, i.e., $a_{\rm here}={32\over3}a_{\rm there}$.} Note that we should actually be a bit more careful. The IR theory has a decoupled $U(1)$ multiplet which gives a contribution $\delta a_{IR}={20\over9}$. As a result, we find that $\delta a\to{28\over9}$ in the large $n$ limit.

This example illustrates our main claim: even though we have a parametrically marginal prepotential deformation, the flow in theory space is not arbitrarily small ($\delta a$ in this case is equivalent to the central charge of seven free hypermultiplets).

Similarly, for flows between the $I_{2,2n-1}$ and $I_{2,2(n-1)-1}$ SCFTs, there is an almost marginal operator, $\hat\CO_{I_{2,2n-1}}$, of dimension $D(\hat\CO_{I_{2,2n-1}})=2\cdot{2n-1\over 2n+1}\sim 2\left(1-{1\over n}\right)$ for $n\gg1$. One finds $\delta a={4(16n^2+32n-5)\over3(4n^2+8n+3)}\ge{172\over45}$ while $\delta a\to{16\over3}$ in the limit of large $n$. Again, we should be careful to include the contribution of a decoupled $U(1)$ multiplet in the IR. This small modification implies that $\delta a\to{28\over9}$ in the large $n$ limit.

We can also consider an $\CN=2\to\CN=1$ breaking RG flow by taking the $I_{3,3n-1}$ ($n\gg1$) theory of \XieJC\ and deforming it by $\delta W=\lambda\CO_{I_{3,3n-1}}$, where $\CO_{I_{3,3n-1}}$ is the $SU(2)_R$ singlet chiral operator of dimension $D={9n-3\over3n+2}\sim3-{3\over n}$. Note that this theory satisfies our consistency condition for parametrically small $\delta a$ in \UVcentral\ since $\epsilon={3\over n}$, $a_{UV}, c_{UV}\sim\CO(\epsilon^{-1})$ (although $d_0\sim\CO(1)$). Indeed, we find that
\eqn\deltaaniiniex{
\delta a={1\over2}\epsilon+\CO(\epsilon^2)\ll1~.
}
Hence, we see that there can be a parametrically close $\CN=1$ fixed point in perturbation theory (we do not see any unitarity bound violations in the chiral sector).

\subsec{An aside on vevs and moduli spaces}
We have assumed that our RG flows involve explicit conformal symmetry breaking in the SCFT and that any vevs we turn on for SCFT operators are sub-leading deformations in the UV (note that if we embed the above AD theories in a parent gauge theory, then we can reinterpret the conformal symmetry breaking as being due to giving a vev to a field coupled to the UV SCFT via an irrelevant interaction \ArgyresJJ). Of course, there are many RG flows in $\CN=2$ SUSY that are driven (or even initiated) by turning on vevs for operators and moving out on a moduli space (we will discuss the case of $\CN=4$ SUSY momentarily). Typically, such deformations are not small from the perspective of the RG flow, because the IR theory must produce a state with dimension one (the dilaton) from a UV theory that usually contains operators of dimensions that are not parametrically close to one.\foot{Note that this concept of the size of the deformation is not related to the existence of a natural metric on the moduli space of the theory.} \foot{Even if we turn on a vev for an operator of dimension $1+\epsilon\sim 1$, this action typically triggers a relevant deformation of the theory that is not small. Heuristically, this is because the operator must acquire its dimension through some interactions, and turning on a vev will result in some highly relevant deformations.} On the other hand, if we turn on vevs for operators that only interact with the SCFT via highly irrelevant interactions, then we may find small deformations from the perspective of the RG flow \AmaritiWC\ (in this paper we do not study such theories).

\newsec{RG flows from $\CN=4$ SYM}
In this section we will study bounds on $\delta a$ for RG flows in which the UV SCFT is the maximally supersymmetric theory. Under the assumption that there are no (as yet undiscovered) non-Lagrangian $\CN=4$ theories (see \refs{\PapadodimasEU,\BeemQXA} for a discussion of this possibility), we can define the maximally supersymmetric theory by the choice of a gauge group, $G$, and a choice of the exactly marginal gauge coupling. The maximal SUSY fixes the superpotential to be
\eqn\NfourW{
W=\sqrt{2}\Tr\left([\Phi_1,\Phi_2]\Phi_3\right)~,
}
where the $\Phi_i$ are $\CN=1$ adjoint chiral superfields. The theory is finite, and the superconformal $R$ charges of the various fields are fixed to be $\tilde R(\Phi_i)={2\over3}$.\foot{Here we write the standard $\CN=1\subset\CN=4$ superconformal $R$ charge that assigns to gauge invariant chiral primaries, $\CO$, dimension $D={3\over2}\tilde R(\CO)$.} The $a$ anomaly can then be easily computed (since we can compute $a$ at any point on the conformal manifold parameterized by the complexified gauge coupling, we can imagine doing this computation in the free field limit of the theory), and we find
\eqn\aNfour{
a_{\CN=4}={8\over3}\cdot|G|~,
}
where $|G|$ is the dimension of the gauge group.

Let us first study deformations that preserve $\CN=4$ SUSY. In this case, the only type of deformation we can make is to go out on the moduli space of the theory. Clearly, the resulting vevs will render at least one of the $\CN=4$ vector multiplets massive. After integrating out this massive matter, we find
\eqn\aiviv{
\delta a\ge{8\over3}~.
}
In particular, $\CN=4$ theories cannot be parametrically close to each other (unless they are just related by a small change in the exactly marginal coupling).\foot{Note that in our conventions, we do not count motion on the conformal manifold as an RG flow (in this case such a motion would correspond to adjusting the exactly marginal gauge coupling of the theory). Such paths result in $\delta a=0$.}

Next let us suppose the RG flow explicitly breaks the UV $\CN=4$ SUSY to $\CN=1,2$. Clearly, we will have to turn on a dimension two operator in order to accomplish this breaking. It is straightforward to check that the corresponding IR theory results in $\delta a$ macroscopically large. For example, suppose we turn on an $\CN=1$-preserving mass term for one of the adjoints
\eqn\deltaW{
W=\sqrt{2}\Tr\left([\Phi_1,\Phi_2]\Phi_3\right)+m\Tr\Phi_3^2~.
}
In the IR, we flow to an $\CN=1$ SCFT with the following superpotential
\eqn\Nonest{
W=h\Tr\left(\left[\Phi_1,\Phi_2\right]^2\right)~.
}
At this fixed point, the superconformal $R$ symmetry assigns $\tilde R_{IR}(\Phi_{1,2})={1\over2}$ and the conformal anomaly is $a_{IR}={9\over4}|G|$, so $\delta a={5\over12}|G|$ cannot be made parametrically small.

\newsec{Conclusion and discussion}
We have seen that, under reasonable assumptions, two $\CN\ge 2$ SCFTs cannot be parametrically close to each other in the space of theories if they are linked by an RG flow initiated by a relevant deformation that preserves $\CN=2$ SUSY (if the UV theory is $\CN=4$, this statement holds even for massive $\CN=1$ deformations). We bound the distance between theories using the change in the $a$ anomaly and the Zamolodchikov metric in the neighborhood of the UV theory.

However, since our analysis of $\CN=2$ theories provides only an existence argument for such bounds (note that our discussion applies equally well to both Lagrangian and non-Lagrangian $\CN=2$ theories), it would be of interest to try to establish an explicit positive bound on $\delta a$ (and $d$). For certain classes of $\CN=2$ theories, one may be able to find an explicit bound on $\delta a$ by using the expression in \ShapereZF\ for the $a$ anomaly in terms of certain geometrical data of the Seiberg-Witten curve (and perhaps bounds on the change in $a$ have an interesting geometrical meaning).

In the appendix, we show that for $\CN=2$ RG flows, the IR superconformal $R$ current can only mix with badly broken symmetries of the short-distance theory under certain assumptions, i.e., that there exists some $\gamma_{\rm univ}>0$ such that the corresponding currents have UV dimension bounded below by $3+\gamma_{\rm univ}$. It would be very interesting to derive a value for $\gamma_{\rm univ}$.

More generally, it might be useful to pursue a \lq\lq bootstrap" program for the RG flow in which we can bound important quantities in different classes of theories just using symmetries and universal principles.

\bigskip\bigskip
\centerline{\bf Acknowledgements}
\smallskip
\noindent
I would like to thank T.~Banks, D.~Gaiotto, Z.~Komargodski, T.~Nishinaka, L.~Rastelli, D.~Simmons-Duffin, Y.~Tachikawa, and M.~\"Unsal for valuable comments and discussions. I am also grateful to the hospitality of San Francisco State University and the Perimeter Institute, where parts of this work were completed. My work is supported by DOE grant DE-FG02-96ER40959.

\appendix{A}{Approximate symmetries in $\CN=2$ SCFTs and the RG flow}
In this appendix, we analyze the role approximately conserved UV currents play in RG flows that emanate from $\CN=2$ SCFTs and are initiated by relevant deformations that break conformal symmetry explicitly. Our discussion applies both to RG flows that preserve the full $\CN=2$ SUSY as well as those that explicitly break $\CN=2\to\CN=1$. We will argue that approximately conserved currents of the $\CN=2$ UV SCFT do not mix with the IR superconformal $R$ current as long as we demand that the UV and IR theories have well-defined Zamolodchikov metrics\foot{In particular, we assume that there are no free non-abelian gauge fields and no free abelian gauge fields with zero coupling.} and that the UV and IR theories can be written in terms of the same degrees of freedom (and are therefore in the same patch of theory space).\foot{Theories with ill-defined Zamolodchikov metrics, or theories in which the UV and IR theories are described in terms of different degrees of freedom necessarily have $d\sim d_0>D_{\rm univ}$ in \dist. If we assume that there exists a non-fine-tuned $a$-function, then there is a $\Delta_{0,\rm univ}>0$ such that $\delta a>\Delta_{0,\rm univ}$.} In particular, this result implies that for the subset of such theories preserving the full $\CN=2$ SUSY

\smallskip
\noindent
$\bullet$ There must be badly broken symmetries in the UV theory that become conserved in the IR and mix with the superconformal $R$ current. The corresponding currents should have UV dimension bounded below by $3+\gamma_{\rm univ}$, where $\gamma_{\rm univ}>0$.

\smallskip
\noindent
Therefore, it follows that

\smallskip
\noindent
$\bullet$ If $d<D_{\rm univ}$ in \dist, then $d\sim d_0$ cannot be parametrically small, i.e., there exists some $d_{\rm univ}>0$ such that $d>d_{\rm univ}$.

\smallskip
\noindent
$\bullet$ If there exists a non-fine-tuned decreasing $a$-function, then there is a $\Delta_{0,\rm univ}>0$ such that $\delta a>\Delta_{0, \rm univ}$.

\smallskip
\noindent
To begin, let us suppose that we have some UV $\CN=2$ SCFT with a set of approximately conserved currents, $\tilde J_{i\mu}$ ($i=1,\cdot\cdot\cdot,N$ is an index for the broken currents; since they are approximately conserved, they have dimensions $3+\epsilon_i$ with $\epsilon_i\ll1$). As discussed in the text, conserved currents may be embedded as descendants in multiplets with real primaries of $SU(2)_R$ spin zero and dimension two (in which case they correspond to $SU(2)_R\times U(1)_R$ currents and are in a multiplet with the stress tensor) or they may be embedded as descendants in multiplets with primaries of $SU(2)_R$ spin one and dimension two (in which case they are \lq\lq flavor" symmetries, i.e., symmetries that commute with the $\CN=2$ superconformal algebra).\foot{Spin one conserved currents can also arise in multiplets that include higher spin conserved currents. In the notation of \DolanZH, these multiplets are denoted $\hat C_{0(j,\bar j)}$ and have dimension $D=2+j+\bar j$ with $j+\bar j>0$ (see the discussion around \stressunit). By the results of \refs{\MaldacenaJN,\AlbaYDA}, these currents must reside in an \lq\lq essentially" free sector of the theory, $\CF$. More precisely, $\CF$ is a sector with an infinite number of higher spin currents whose correlation functions are equivalent to those of a free theory. Since we know that these spin one currents do not mix with the superconformal $R$ current in a free theory (they do not enter into $a$-maximization), it then seems reasonable to assume that these currents can never mix with the superconformal $R$ current. Therefore, even if such symmetries emerge in the IR, we ignore them below.} When symmetries are broken, the corresponding currents pair up with some of the symmetry-breaking data to form long multiplets.

Let us first focus on approximate symmetries that arise from real primaries of $SU(2)_R$ spin zero, $J_I$ (where $I=1,\cdot\cdot\cdot, N_S$). If the corresponding currents are approximately conserved, it is reasonable to assume that---at least in one duality frame---the lack of conservation arises from weakly coupling two or more decoupled SCFTs, $\CT_I$ (with $I$ counting the decoupled sectors; note that the $\CT_I$ may be free or interacting) so that the $N_S$ independent $SU(2)_R\times U(1)_R$ currents and stress tensors are broken to a single set of $R$ symmetry currents and stress tensor for the coupled system.\foot{For example, consider the $SU(3)$ $\CN=2$ gauge theory with six hypermultiplets at very strong coupling. This theory has approximately conserved symmetries arising from several $J_I$ (in addition to an exact $SU(2)_R\times U(1)_R$ symmetry) even though it naively appears to have a single sector. In fact, as explained in \ArgyresCN, this theory secretly has several modules since it is dual to a copy of the $E_6$ theory \MinahanFG\ with a weakly gauged $SU(2)\subset E_6$ subgroup coupled to two doublets.}

If we assume that the coupling between the $\CT_I$ is due to an exactly marginal deformation, it is straightforward to check that this coupling occurs via an $\CN=2$ gauging.\foot{One amusing but rather indirect way to understand this statement is the following. Suppose the $\CT_I$ are not coupled by an $\CN=2$ gauging. Then, we would be forced to couple the $\CT_I$ by introducing terms of the form $\delta W=\lambda\CO$ with $\CO=\CO_1\cdot\cdot\cdot\CO_{\hat N}$ a composite operator built out of gauge-invariant ($\CN=1\subset\CN=2$) primaries from the various $\CT_I$. Since our theory is an SCFT, it must be the case that $\CO$ has $SU(2)_R$ spin one (with $I_3\subset SU(2)_R$ charge $+1$) and $U(1)_R$ charge $+2$ (these charges are with respect to the symmetries of the combined SCFT).

Since the $\CT_I$ are SCFTs and we couple them via an exactly marginal interaction, each $\CO_a$ must have either
\smallskip
\noindent
{\bf (i)} $SU(2)_R$ spin zero and $U(1)_R$ charge $R\ge2$ (and therefore dimension $D(\CO_a)={R\over2}$ and $J_I(\CO_a)=R$, where $J_I=R_{\CN=2}-2I_3$ for the SCFT $\CT_I$ to which $\CO_a$ belongs).

\smallskip
\noindent
{\bf (ii)} $SU(2)_R$ spin $j_R\ge{1\over2}$ and $U(1)_R$ charge zero (and therefore dimension $D(\CO_a)=2j_R$ and $J_I(\CO_a)=-2j_R$).

\smallskip
\noindent
{\bf (iii)} $SU(2)_R$ spin $j_R\ge{1\over2}$ and $U(1)_R$ charge $R\ge2$ (and therefore dimension $D(\CO_a)=2j_R+{R\over2}$ and $J_I(\CO_a)=R-2j_R$).
\smallskip

Since $\CO$ has $R=2$, it must be the case that there is one and only one $\CO_a$ of type {\bf(i)} or {\bf(iii)}, and this $\CO_a$ must have $R=2$. Let us suppose it is of type {\bf(i)}. In this case, $\CO_a$ is just the lowest component of a free $U(1)$  $\CN=2$ vector superfield. As a result, $\CO$ must in fact be marginally irrelevant, since adding it to the superpotential necessarily breaks the symmetry that rotates $\CO_a$ by a phase \GreenDA. Therefore, such operators cannot contribute to the deformation coupling the two theories. Next let us suppose that our $\CO_a$ is of type {\bf (iii)}. In this case, $\CO_a$ is either $SU(2)_R$ spin one or one half. If it is $SU(2)_R$ spin one, then $\CO=\CO_a$, and there is no coupling between the $\CT_I$ unless we also have an $\CO$ with $\CO_a$ of $SU(2)_R$ spin one half. For this operator, we must have that $\CO=\CO_a\CO'$, where $\CO'$ is an operator from a different SCFT of type {\bf(ii)} with $j_R={1\over2}$ and $R=0$. As a result, we see that $\CO'$ is part of a free hypermultiplet, and so $\CO$ must be marginally irrelevant since adding it to the superpotential breaks the symmetry that rotates $\CO'$ by a phase \GreenDA. It then follows that if we wish to weakly couple the $\CT_I$ via an exactly marginal deformation, we must do so by an exactly marginal gauging.} Note that we could also attempt to couple the $\CT_I$ by a relevant deformation. However, relevant prepotential couplings between the theories are prohibited by unitarity, and moment map couplings simply reduce to mass terms for free hypermultiplets.

As a result, in $\CN=2$ SCFTs with approximately conserved $J_I$, we expect to have terms of the form
\eqn\Wcoupl{
\delta W=\sum_i\left({\Theta_i\over2\pi}+{4\pi i\over g_i^2}\right)\Tr\ W_i^2+\sqrt{2}\sum_{i,I}\Tr\ \Phi_i \cdot\mu_I^i+\cdot\cdot\cdot~,
}
where $\mu_I^i$ is a holomorphic moment map (with $j_{I, R}=1$ and $R_I=0$) for some symmetry, $G_i$, in $\CT_I$ that we weakly gauge ($\Phi_i$ is the corresponding chiral adjoint superfield, $(\Theta_i, g_i)$ are the corresponding theta angles and gauge couplings, and $i=1,\cdot\cdot\cdot, N_G$ runs over the global symmetry groups we gauge). If $G_i$ is not a symmetry of some particular $\CT_I$, then we set $\mu_I^i=0$ in \Wcoupl. Since $J_I(\mu_J^i)=-2\delta_{IJ}$ (as long as $G_i$ is a symmetry of $\CT_I$) and $J_I(\Phi_i)=0$, it is clear that \Wcoupl\ breaks the various $J_I$ symmetries (and the $J_{\Phi_i}$ symmetries which assign $J_{\Phi_i}(\Phi_j)=2\delta_{ij}$ and $J_{\Phi_i}(\mu_I^j)=0$) down to a diagonal symmetry, $J$, which satisfies
\eqn\Jdiag{
J(\mu_I^i)=-J(\Phi_i)=-2~.
}

Note that each individual $J_I$ is anomalous as well since, by unitarity, if $G_i$ is a symmetry of $\CT_I$,\foot{By construction, there must be at least one non-trivial $G_i$ for each $\CT_I$.} then $\Tr J_IG_iG_i=\Tr R_{\CN=2, I}G_iG_i=-\tau_{G_i}|_{\CT_I}<0$ (here $\tau_{G_i}|_{\CT_I}$ is the two point function of the gauge currents, $J_{G_i}$, in $\CT_I$). Furthermore, note that $\Tr J_{\Phi_i}G_iG_i=2C_2({\rm Adj}_i)>0$, where $C_2({\rm Adj}_i)$ is the quadratic Casimir of the adjoint representation of $J_G$. Anomaly freedom of $J$ and marginality of the gauge couplings require
\eqn\margtau{
2C_2({\rm Adj}_i)=\sum_I\tau_{G_i}|_{\CT_I}~, \ \ \ \sum_I\tau_{G_iG_{j\ne i}}|_{\CT_I}=0~.
}
Neglecting mixed anomalies, we can form $N_S$ non-anomalous combinations out of the $N_G+N_S$ $\left\{J_{\Phi_i}, J_I\right\}$ operators (one of which is $J$ in \Jdiag\ and the rest of which can be parameterized by $a^{\Phi_i}J_{\Phi_i}+\sum a^IJ_I$) by solving the following $N_G$ equations (the index $i$ is not summed over)
\eqn\anomfree{
2a^{\Phi_i}C_2({\rm Adj}_i)-\sum_I a^I\tau_{G_i}|_{\CT_I}=0~.
}
All of these currents except for $J$ are broken by the superpotential terms \Wcoupl\ (any mixed anomalies will further reduce the set of available non-anomalous symmetries).

Given our above $\CN=2$ theory with approximately conserved $J_I$ and $J_{\Phi_i}$ symmetries, we can now consider turning on a relevant deformation. Let us suppose that we preserve $\CN=2$ SUSY. Since we assume that the IR theory has a well-defined Zamolodchikov metric and that it can be written in terms of the same degrees of freedom as the UV theory (in particular, that the non-abelian symmetries of the UV theory do not decouple in the IR), then \Wcoupl\ must still be present in the IR. Since the $\mu_I^i$ and $\Phi_i$ in \Wcoupl\ cannot be charged under (broken or conserved) flavor symmetries (otherwise these symmetries would not commute with the gauge symmetry), it follows that the approximately conserved $J_I$ and $J_{\Phi_i}$ cannot mix with the IR superconformal $R$ current (they would result in $\tilde R_{IR}(\delta W)\ne2$).

Note that we have not yet discussed emergent flavor symmetries. In fact, such symmetries turn out to be irrelevant for our purposes since they are non-chiral and do not mix with the IR superconformal $R$ current, $\tilde R_{IR}$. Furthermore, there cannot be emergent symmetries that are linear combinations, $\hat J_A$, of broken flavor symmetries with the approximately conserved $J_I$ or $J_{\Phi_i}$ symmetries, since some of the $\Phi_i\mu^i_I$ would necessarily be charged under these linear combinations.

Next let us consider RG flows with $\CN=2\to\CN=1$ breaking. In this case, the IR $\Phi_i\mu^i_I$ couplings in \Wcoupl\ need not be related to IR gauge couplings (since the long-distance theory only has $\CN=1$ SUSY). One might then imagine that these superpotential terms could be irrelevant in the IR and flow to zero (without the Zamolodchikov metric being ill-defined), thus allowing some non-anomalous combinations of the $J_I$ to mix with $\tilde R_{IR}$. In fact, this is not possible. To understand this statement, first note that \margtau\ and \anomfree\ imply 
\eqn\eqncons{
\sum_I(a^{\Phi_i}-a^I)\tau_{G_i}|_{\CT_I}=0~.
}
Defining $\rho^{Ii}=a^{\Phi_i}-a^I$, we see that the corresponding approximate UV $J'$ symmetry assigns charge $J'(\Tr\ \Phi_i\cdot\mu^i_I)=2\rho^{Ii}$ (when $G_i$ is a symmetry of $\CT_I$). In particular,  there must be at least one $\rho^{I_0i_0}<0$ and another $\rho^{I_1i_0}>0$ (since the $\tau_{G_i}|_{\CT_I}>0$ by unitarity). Now, suppose that $J'$ mixes with $\tilde R_{IR}$. Then, the corresponding $\Tr\ \Phi_{i_0}\cdot\mu^{i_0}_{I_{0,1}}$ operators would have to be irrelevant in the IR since they would have $\tilde R_{IR}$ charge different from two. However, if these operators are irrelevant, then they are primaries of the IR theory. It then follows that one of these two operators will be relevant in the IR (since their IR dimensions are then given by $D_{IR}={3\over2}\tilde R_{IR}$). This is a contradiction.

Now let us consider flavor symmetries in RG flows with $\CN=2\to\CN=1$ breaking. Again, if an approximately conserved $\CN=2$ flavor symmetry flows to some non-chiral symmetry in the IR, then we do not expect that it should contribute to the IR superconformal $R$ current. On the other hand, we can imagine that the approximately conserved UV $\CN=2$ flavor symmetry flows to a chiral symmetry in the IR (since the IR theory is only $\CN=1$). All such examples we are aware of are just symmetries of the form corresponding to $\hat J_A$ discussed in a previous paragraph. As a result, we again do not expect approximately conserved UV flavor symmetries to mix with the IR superconformal $R$ current for the $\CN=1$ fixed point.

\listrefs
\end